\documentclass[paper,11pt]{JHEP}
\usepackage[centertags]{amsmath}
\usepackage{amsfonts} \usepackage{amssymb} \usepackage{amsthm}
\usepackage{graphicx}
\newcommand{\eq}[1]{Eq. (\ref{#1})}

\newcommand{\ket}[1]{|{#1}\rangle}
\newcommand{\braket}[2]{\langle{#1}|{#2}\rangle}
\newcommand{\ii}{\mathrm{i}}
\newcommand{\ee}{\mathrm{e}}
\newcommand{\one}{{\rm 1\kern -.9mm l}}
\newcommand{\uno}{\mbox{1\!\negmedspace1}}

\newcommand{\Tr}{\mathrm{Tr}\,}
\newcommand{\tr}{\mathrm{tr}\,}

\newcommand{\smod}{S_{\rm mod}(q,\tilde q;{\cal M}_{k})}
\newcommand{\smodh}{S_{\rm mod}(q,\tilde q;{\widehat{\cal M}_{k}})}
\newcommand{\Gammab}{\boldsymbol{\Gamma}}

\newcommand{\Lhol}{\Lambda_{\scriptscriptstyle\mathrm{hol}}}
\newcommand{\Lholt}{\widehat{\Lambda}_{\scriptscriptstyle\mathrm{hol}}}
\newcommand{\Decol}{\Delta_{\scriptscriptstyle\mathrm{color}}}
\newcommand{\Deflav}{\Delta_{\scriptscriptstyle\mathrm{flavor}}}
\newcommand{\Decolt}{\widetilde{\Delta}_{\scriptscriptstyle\mathrm{color}}}
\newcommand{\Deflavt}{\widetilde{\Delta}_{\scriptscriptstyle\mathrm{flavor}}}
\newcommand{\ssb}[2]{{\scriptstyle{#1 \brack #2}}}
\title{Instanton effects in $\mathcal{N}=1$ brane models and the K\"ahler
metric of twisted matter
}
\author{Marco Bill\`o, Marialuisa Frau, Igor Pesando\\
Dipartimento di Fisica Teorica, Universit\`a di Torino\\
and I.N.F.N. - sezione di Torino \\
Via P. Giuria 1, I-10125 Torino, Italy
}
\author{Paolo Di Vecchia\\
The Niels Bohr Institute, Blegdamsvej 17, DK-2100 Copenhagen \O{},
Denmark\\
and NORDITA, Roslagstullsbacken 23, SE-10691 Stockholm, Sweden
}
\author{Alberto Lerda\\
Dipartimento di Scienze e Tecnologie Avanzate, Universit\`a del Piemonte Orientale\\
and I.N.F.N. - Gruppo Collegato di Alessandria - sezione di Torino\\
Via V. Bellini 25/G, I-15100 Alessandria, Italy
}
\author{Raffaele Marotta\\
I.N.F.N. - sezione di Napoli
and Dipartimento di Scienze Fisiche, Universit\`a di Napoli.\\
Complesso Universitario Monte S. Angelo - ed. G\\
Via Cintia - I-80126 Napoli, Italy}
\abstract{We consider locally consistent systems of magnetized D9
branes on an orbifolded six-torus which support $\mathcal{N}=1$ gauge
theories. In such realizations, the matter multiplets arise from ``twisted''
strings connecting different stacks of branes. The introduction of Euclidean
5 branes (E5) wrapped on the six-dimensional compact space  leads to
instanton effects. For instance, if the system is engineered
so as to yield SQCD, a single E5 brane may account for the ADS/TVY superpotential.
We discuss the subtle interplay that exists between the annuli diagrams
with an E5 boundary and the holomorphicity properties  of the effective
low-energy action of the $\mathcal{N}=1$ theory.
The consistency of this picture allows to obtain information on
the K\"ahler metric of the chiral matter multiplets arising from
twisted strings.
}
\keywords{Superstrings, D-branes, Gauge Theories, Instantons}
\preprint{DFTT/15/2007\\NORDITA-2007-26}

\begin{document}

\section{Introduction}
\label{sec:intro}
While non-perturbative instanton  effects have been analyzed in great detail in
field theory and can be evaluated by means of complete and clear algorithms (for
reviews, see for instance Refs. \cite{Dorey:2002ik,Bianchi:2007ft}), the study
of these effects in string theory is still at an early stage and, despite some
remarkable progresses in the last few years, further work is still needed to
reach a similar degree of accuracy in their computation. This would be very
important not only for including string corrections to the effects that have
been already computed with field theoretical methods, but especially to derive
new non-perturbative effects of purely stringy origin that could play a relevant
role in the applications of string theory to phenomenology. Recently this
possibility has been intensively investigated from several different points of
view and has received considerable
attention~\cite{Beasley:2005iu}--\cite{Blumenhagen:2007bn}.

However, in order to learn how to deal with non-perturbative effects in string
theory and gain a good control on the results, it is very important also to
reproduce, using string methods, the non-perturbative effects already known from
field theory. To this aim, toroidal orbifolds of Type II string theory (for a
review see Ref. \cite{Blumenhagen:2006ci}) are very useful since they provide a
concrete framework in which one can perform explicit calculations of instanton
effects. For example, they can be used to engineer ${\cal{N}}=2$ super
Yang-Mills (SYM) theories and study the instanton induced prepotential, as
discussed in detail in Ref. \cite{Billo:2006jm}. In a recent paper
\cite{Billo:2007sw} we have extended this procedure by compactifying six
dimensions on $(\mathcal{T}_2^{(1)}\times\mathcal{T}_2^{(2)})/\mathbb{Z}_2\,
\times \mathcal{T}_2^{(3)}$ and by including the contribution of the mixed
annuli diagrams, as advocated in Refs.
\cite{Blumenhagen:2006xt,Akerblom:2006hx,Akerblom:2007uc}. In particular we have
shown that the non-holomorphic terms in these annulus amplitudes precisely
reconstruct the appropriate K{\"{a}}hler metric factors that are needed to write
the instanton correlators in terms of purely holomorphic variables. In this way
the correct holomorphic structure of the instanton induced low energy effective
action in the Coulomb branch of the $\mathcal{N}=2$ SYM theory has been
obtained.

In the present paper we apply this procedure to ${\cal{N}}=1$ SYM theories that
we engineer by means of stacks of magnetized fractional D$9$ branes  in a
background given by the product of  $\mathbb{R}^{1,3}$ times a six-dimensional
orbifold $(\mathcal{T}_2^{(1)}\times\mathcal{T}_2^{(2)}\times\mathcal{
T}_2^{(3)})/(\mathbb{Z}_2\times \mathbb{Z}_2)$. A single stack of fractional D9
branes, that we call ``color'' branes, supports on its world-volume a pure
${\cal{N}}=1$ gauge theory. Matter chiral multiplets can be obtained by
introducing a second stack of magnetized fractional D$9$ branes, called
``flavor'' branes, that belong in general to a different irreducible
representation of the orbifold group, and by considering the massless open
strings having one endpoint on the color branes and the other on the flavor
branes. In this framework one can also engineer ${\cal{N}}=1$ super QCD by
suitably introducing a third stack of magnetized fractional D$9$ branes, in such
a way that the massless open strings connecting the color branes and the two
types of flavor branes correspond respectively to the right and left-handed
quarks and their super-partners, and hence give rise to a vector-like theory as
described in Section. \ref{sec:model}.

To study instanton effects in this set-up one has to add a stack of fractional
Euclidean D5 branes (E5 branes for short) that completely wrap the internal
manifold and hence describe point-like configurations from the four-dimensional
point of view. If the wrapping numbers and magnetization of these E5 branes are
the same as those of the color D9 branes, one has a stringy realization of
ordinary gauge theory instantons%
\footnote{In fact, these D9/E5 systems are essentially a
T-dual version of the D3/D(--1) systems which, in un-compactified set-ups, are
well-known to realize at the string theory level the gauge instantons and their
moduli, described \`a la ADHM \cite{Witten:1995im}-\cite{Billo:2002hm}.}.
If instead their wrapping numbers and magnetization are different
from the color branes, one obtains ``exhotic'' instanton
configurations of purely stringy nature. In this paper we will not
explicitly consider this possibility, even if our methods
could be used also in this case. On the contrary, following
the procedure outlined in Refs. \cite{Billo:2002hm,Billo:2006jm}, we compute
using string methods the superpotential in $\mathcal{N}=1$ SYM theories induced by
gauge instantons.
In doing so, the contribution of mixed annulus diagrams with a
boundary attached to the E5 branes, which
are of the same order in the string coupling constant as the disk diagrams which
account for the moduli measure, has to
be taken into account.

As noticed in the literature
\cite{Blumenhagen:2006xt,Abel:2006yk,Akerblom:2006hx,Akerblom:2007uc}, in
supersymmetric situations these mixed annulus amplitudes are related in a precise way to
the 1-loop corrections to the gauge coupling constant of the color gauge
theory and the physical origin of this
identification has been discussed in Ref. \cite{Billo:2007sw}.
This relation can then be used to compare the explicit
expression of the mixed annulus amplitudes to the general
formula \cite{Dixon:1990pc,Kaplunovsky:1994fg,Louis:1996ya} that expresses the
1-loop corrections to the gauge coupling computed in string theory in terms of
the fields and geometrical quantities that appear in the effective supergravity
theory, such as the K\"ahler metrics for the various multiplets.
Exploiting this fact, we explicitly compute the mixed annulus
diagrams in our orbifold models and extract from them information on the
K{\"{a}}hler metric for the matter
multiplets. We then perform two checks on our results.

First, we consider the 1-instanton induced superpotential in the set-up
corresponding to $\mathcal{N}=1$ SQCD. In Refs. \cite{Akerblom:2006hx,Argurio:2007vq}
it has already been shown that the stringy instanton calculus in this case
reproduces the ADS/TVY superpotential \cite{Affleck:1983mk}
(see also Ref. \cite{Taylor:1982bp}). Here
we discuss in detail the r\^ole of the mixed annuli contributions and show that
they are crucial in making this superpotential holomorphic when expressed in terms
of the variables appropriate to the low-energy supergravity description.

Second, we exploit the fact that the K\"ahler metrics of the matter multiplets
enter crucially in the relation between the holomorphic superpotential couplings
in the effective Lagrangian and the physical Yukawa couplings for
the canonically normalized fields. We consider the expression of the latter
provided in Ref. \cite{Cremades:2004wa} for the field-theory limit
of magnetized brane models, and show that, after transforming it to the
supergravity basis, it becomes purely holomorphic.

The paper is organized as follows. In Section \ref{sec:model} we describe the
set-up we utilize for realizing ${\cal{N}}=1$ supersymmetric gauge theories.
Section \ref{sec:inst_branes} is devoted to the description of the instanton
calculus in this set-up. In Section \ref{subsec:mix_ann} we compute the mixed
annulus diagrams while in Section \ref{sec:relation}
we discuss the relation with the K{\"{a}}hler metric for the matter
fields; furthermore we check the holomorphicity of the 1-instanton induced
superpotential. In the last section we show that our expressions yield
holomorphic cubic superpotential couplings of the matter multiplets
if we start from the physical Yukawa couplings in
magnetized brane models computed in  Ref. \cite{Cremades:2004wa}.
Finally, many technical details are given in the Appendix.

\section{Local $\mathcal{N}=1$ brane models with chiral matter}
\label{sec:model}
A way to realize a $\mathcal{N}=1$ SYM theory is to place a stack of
fractional D$9$ branes in a background given by
the product of $\mathbb{R}^{1,3}$ times a six-dimensional orbifold
\begin{equation}
\frac{\mathcal{T}_2^{(1)}\times\mathcal{T}_2^{(2)}\times\mathcal{
T}_2^{(3)}}{\mathbb{Z}_2\times \mathbb{Z}_2}~.
\label{orbifold}
\end{equation}
For each torus $\mathcal{T}_2^{(i)}$, the string frame metric
and the $B$-field%
\footnote{Without loss of generality, in the following we will actually set the
$B$-field to zero.}
are parameterized by the K\"ahler and complex structure
moduli, $T^{(i)}=T_1^{(i)}+\ii \,T_2^{(i)}$ and $U^{(i)}=U_1^{(i)}+
\ii \,U_2^{(i)}$ respectively.
For our precise conventions we refer to Appendix
\ref{appsub:geom}. The ten-dimensional string coordinates $X^M$ and $\psi^M$ are
split as
\begin{equation}
X^M \to (X^\mu, Z^i)
~~~~{\rm and}~~~~
\psi^M \to (\psi^\mu, \Psi^i)~,
\label{coordinates}
\end{equation}
where $\mu=0,1,2,3$ and the complex coordinates $Z^i$ and $\Psi^i$, defined in
Eq. (\ref{zipsii}), are orthonormal in the metric of the $i$-th torus.
Also the (anti-chiral) spin-fields $S^{\dot{\mathcal{A}}}$ of the RNS
formalism in ten dimensions factorize in a product of
four-dimensional and internal spin-fields, and the precise splitting is
given in Eq. (\ref{spin}).
The $\mathbb{Z}_2\times \mathbb{Z}_2$ orbifold group in (\ref{orbifold})
contains three non-trivial elements $h_i$ ($i=1,2,3$). The element $h_i$ leaves
the $i$-th torus $\mathcal{T}_2^{(i)}$ invariant
while acting as a reflection on the remaining
two tori.

The above geometry can also be described in the so-called supergravity basis
using the complex
moduli $s$, $t^{(i)}$ and $u^{(i)}$, whose relation with the previously
introduced quantities in the string basis is
\cite{Lust:2004cx,Blumenhagen:2006ci}
\begin{equation}
\begin{aligned}
&\mathrm{Im}(s) \equiv s_2 = \frac{1}{4\pi}\,\ee^{-\phi_{10}}\,
T_2^{(1)}T_2^{(2)}T_2^{(3)}~,
\\
&\mathrm{Im}(t^{(i)}) \equiv t_2^{(i)} = \ee^{-\phi_{10}} T_2^{(i)}~,
\\
& u^{(i)} = u_1^{(i)} + \ii\, u_2^{(i)} = U^{(i)}~,
\end{aligned}
\label{stu}
\end{equation}
where $\phi_{10}$ is the ten-dimensional dilaton. The real parts of $s$ and
$t^{(i)}$
are related to suitable RR potentials.
In terms of these variables, the $\mathcal{N}=1$ bulk K\"ahler potential is
given by
\cite{Antoniadis:1996vw}
\begin{equation}
K = -\log (s_2) -\sum_{i=1}^3 \log(t_2^{(i)}) -
\sum_{i=1}^3 \log(u_2^{(i)})~.
\label{kpot}
\end{equation}

\paragraph{Colored and flavored branes}
In this orbifold background we place a stack of $N_a$ fractional
D$9$ branes (hereinafter called colored branes and labeled by an index $a$)
which for definiteness are taken to transform in the trivial irreducible
representation $R_0$ of the orbifold group.
The massless excitations of the open strings attached to these branes fill the
$\mathcal{N}=1$
vector multiplet in the adjoint representation of $\mathrm{U}(N_a)$.
The disk interactions of the corresponding vertex
operators reproduce, in the field theory limit $\alpha'\to 0$,
the $\mathcal{N}=1$ SYM action with $\mathrm{U}(N_a)$ gauge group, which in the
Euclidean signature appropriate to discuss instanton effects, reads
\begin{equation}
S_{\rm SYM}=\frac{1}{g_a^2}\,
\int d^4x ~{\rm
Tr}\,\Big\{\frac{1}{2}\,F_{\mu\nu}^2
-2\,\bar\Lambda_{\dot\alpha}\bar D\!\!\!\!/^{\,\dot\alpha \beta}
\Lambda_\beta \Big\}~,
\label{n1}
\end{equation}
where the tree-level Yang-Mills coupling constant $g_a$ is given by
\begin{equation}
\frac{1}{g_a^2} =
\frac{1}{4\pi}\,\ee^{-\phi_{10}}\,T_2^{(1)}T_2^{(2)}T_2^{(3)}
= s_2~.
\label{gym}
\end{equation}

Richer models can be found if we introduce additional stacks of
fractional D9-branes, distinguished with a subscript $b$, that belong to
various irreducible representations of the orbifold group
and can be magnetized. In general, we will have $N_b$ branes of
type $b$,
which we will call flavor branes, and $n_b^{(i)}$ will be their wrapping
number around the $i$-th torus.
These branes admit a constant magnetic field on the $i$-th torus
\begin{equation}
F_b^{(i)}= f_b^{(i)}\,dX^{2i+2}\wedge dX^{2i+3} =
\ii\,\frac{f_b^{(i)}}{T_2^{(i)}}\,
dZ^i\wedge d{\bar Z}^i ~.
\label{fi}
\end{equation}
The generalized Dirac quantization condition requires that the
first Chern class $c_1(F_b^{(i)})$ be an integer, which, in our conventions,
implies that
\begin{equation}
2\pi\alpha' f_b^{(i)} = \frac{m_b^{(i)}}{n_b^{(i)}}
\label{nm}
\end{equation}
with $m_b^{(i)}\in \mathbb{Z}$.
In terms of the angular parameters $\nu_b^{(i)}$, defined by
\begin{equation}
2\pi\alpha' \frac{f_b^{(i)}}{T_2^{(i)}} = \tan \pi\nu_b^{(i)}~~~~{\rm
with}~~~~0\leq \nu_b^{(i)}
< 1~,
\label{nui}
\end{equation}
it is possible to show that bulk $\mathcal{N}=1$ supersymmetry is preserved
if%
\footnote{Other possibilities are
$-\nu_b^{(1)}-\nu_b^{(2)}+\nu_b^{(3)}=0$;
$-\nu_b^{(1)}+\nu_b^{(2)}-\nu_b^{(3)}=0$;
$\nu_b^{(1)}+\nu_b^{(2)}+\nu_b^{(3)}=2$. They are all related to the
position (\ref{nu123}) by obvious changes.}
\begin{equation}
\nu_b^{(1)}-\nu_b^{(2)}-\nu_b^{(3)}=0~.
\label{nu123}
\end{equation}
The presence of the magnetic fluxes implies that the open strings stretching
between two different types of branes ({\it e.g.} the D$9_b$/D$9_a$ strings)
are twisted. This means that the internal string coordinates $Z^i$ and $\Psi^i$
have the following twisted monodromy properties
\begin{equation}
Z^i\big({\rm e}^{2\pi{\rm i}} z\big)= \,{\rm e}^{2\pi{\rm
i}\nu^{(i)}_{b}}\,Z^i(z)~~~{\mbox{and}}~~~
\Psi^i\big({\rm e}^{2\pi{\rm i}} z\big)= \eta\,{\rm e}^{2\pi{\rm
i}\nu^{(i)}_{b}}\,\Psi^i(z)~,
\label{monodromy}
\end{equation}
where $\eta=+1$ for the NS sector and $\eta=-1$ for the R sector.
If also the color branes are magnetized, we have to replace in
(\ref{nu123}) and (\ref{monodromy})
$\nu^{(i)}_{b}$ with
$\nu^{(i)}_{ba}=\nu^{(i)}_{b}-\nu^{(i)}_{a}$, which describe
the relative magnetization of the two stacks of branes.
When no confusion is possible, we will denote the twist angles simply by
$\nu^{(i)}$.

As is well-known, in a toroidal orbifold
compactification with wrapped branes there are unphysical closed
string tadpoles that must be canceled to have a globally
consistent model. Usually this cancellation is achieved by
introducing an orientifold projection and suitable orientifold
planes. Like in other cases treated in the literature, in this
paper we take a ``local'' point of view and assume that the brane
systems we consider can be made fully consistent with an orientifold
projection.

\begin{figure}
\begin{center}
\begin{picture}(0,0)%
\includegraphics{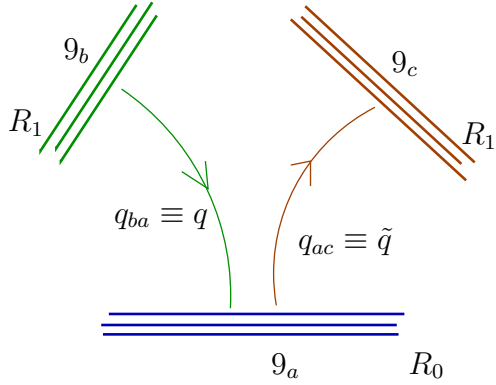}%
\end{picture}%
\setlength{\unitlength}{2171sp}%
\begingroup\makeatletter\ifx\SetFigFont\undefined%
\gdef\SetFigFont#1#2#3#4#5{%
  \reset@font\fontsize{#1}{#2pt}%
  \fontfamily{#3}\fontseries{#4}\fontshape{#5}%
  \selectfont}%
\fi\endgroup%
\begin{picture}(6175,4386)(276,-3844)
\put(1801,-136){\makebox(0,0)[rb]{\smash{{\SetFigFont{12}{14.4}
{\familydefault}{\mddefault}{\updefault}$9_b$}}}}
\put(5551,-286){\makebox(0,0)[rb]{\smash{{\SetFigFont{12}{14.4}
{\familydefault}{\mddefault}{\updefault}$9_c$}}}}
\put(3151,-2011){\makebox(0,0)[rb]{\smash{{\SetFigFont{12}{14.4}
{\familydefault}{\mddefault}{\updefault}$q_{ba}\equiv q$}}}}
\put(5251,-2311){\makebox(0,0)[rb]{\smash{{\SetFigFont{12}{14.4}
{\familydefault}{\mddefault}{\updefault}$q_{ac}\equiv \tilde{q}$}}}}
\put(4201,-3736){\makebox(0,0)[rb]{\smash{{\SetFigFont{12}{14.4}
{\familydefault}{\mddefault}{\updefault}$9_a$}}}}
\put(1276,-961){\makebox(0,0)[rb]{\smash{{\SetFigFont{12}{14.4}
{\familydefault}{\mddefault}{\updefault}$R_1$}}}}
\put(6451,-1111){\makebox(0,0)[rb]{\smash{{\SetFigFont{12}{14.4}
{\familydefault}{\mddefault}{\updefault}$R_1$}}}}
\put(5851,-3736){\makebox(0,0)[rb]{\smash{{\SetFigFont{12}{14.4}
{\familydefault}{\mddefault}{\updefault}$R_0$}}}}
\end{picture}%
\end{center}
\caption{Schematization of the brane system we consider, and of its spectrum of
 chiral multiplets; see the text for more details.}
\label{fig:9a9b9c}
\end{figure}

\paragraph{$\mathcal{N}=1$ SQCD with magnetized branes}
In the following, we will be mostly interested in studying instanton
effects in $\mathcal{N}=1$ SQCD with $N_F$ flavors. In our orbifold background
we can realize this model by taking two stacks of
flavored fractional D9 branes, denoted by $b$ and $c$ respectively, both
belonging to a different representation of the orbifold
group with respect to the color branes; see Figure \ref{fig:9a9b9c} for a
pictorial representation of the system we consider.
For definiteness, we take the
$R_1$ representation as defined in Appendix \ref{appsub:geom}.

If the twist angles satisfy the $\mathcal{N}=1$ supersymmetry condition
$\nu_{ba}^{(1)}-\nu_{ba}^{(2)}-\nu_{ba}^{(3)}=0$, the massless states
of the D$9_b$/D$9_a$ strings fill up a chiral multiplet $q_{ba}\equiv
q$, which transforms in the anti-fundamental representation $\bar{N}_a$
of the color group and appears with a flavor degeneracy
\begin{equation}
 \label{Nab}
N_b |I_{ab}|~,
\end{equation}
where $I_{ab}$ is the number of Landau levels for the $(a,b)$
``intersection'', namely
\begin{equation}
I_{ab} = \prod_{i=1}^3\big(m_a^{(i)}n_b^{(i)}-
m_b^{(i)}n_a^{(i)}\big)=-I_{ba}~.
\label{iab}
\end{equation}
The complex scalar, denoted with an abuse of notation by the
same letter $q$ used for the whole multiplet, arises from
the NS sector and is described by the vertex operator (\ref{vertn1scal}).
Its supersymmetric partner is a chiral fermion $\chi_\alpha$ described
by the vertex operator (\ref{vertn1ferm}) of the R sector, which
is connected to the scalar vertex by the $\mathcal{N}=1$ supersymmetry generated
by the
open string supercharges preserved by the $\mathbb{Z}_2\times\mathbb{Z}_2$
orbifold.

In an analogous way, we can analyze the open strings stretching
between the color branes and the flavor branes of type $c$.
If the twist angles are such that
$\nu_{ac}^{(1)}-\nu_{ac}^{(2)}-\nu_{ac}^{(3)}=0$, then the massless states
of the D$9_a$/D$9_c$ strings (notice the orientation!) fill up a chiral
multiplet $q_{ac}\equiv \tilde q$
that transforms in the fundamental representation ${N}_a$
of the color group and appears with a flavor degeneracy
\begin{equation}
 \label{Nac}
N_c |I_{ac}|~,
\end{equation}
where $I_{ac}$ is the number of Landau levels for the $(a,c)$
``intersection''. The bosonic and fermionic components of the
multiplet $\tilde q$ are described, respectively, by the vertex operators
(\ref{vertn1scalt}) and (\ref{vertn1fermt}) which
are also related to each other by the $\mathcal{N}=1$ supersymmetry
preserved by the orbifold.

This set-up provides a realization of $\mathcal{N}=1$ SQCD if we
arrange the branes in such a way that the flavor degeneracies
(\ref{Nab}) and (\ref{Nac}) are equal:
\begin{equation}
N_b |I_{ab}|=N_c |I_{ac}|\equiv N_F~.
\label{nf}
\end{equation}
In this way we engineer the same number
$N_F$ of fundamental and anti-fundamental chiral multiplets, which
will be denoted by $q_f$ and ${\tilde q}^f$ with $f=1,\ldots,N_F$.

The field-theory limit of the disk amplitudes involving the fields of the chiral
multiplets and those of the vector
multiplet yields the $\mathcal{N}=1$ SQCD action; for instance, the kinetic term
of
the scalars arises in the form
\begin{equation}
\int d^4x ~\sum_{f=1}^{N_F}\Big\{D_\mu q^{\dagger f} \,D^\mu {q}_f
 + D_\mu {\tilde q}^f\,D^\mu \tilde{q}^\dagger_f\Big\}~,
\label{kinq}
\end{equation}
where we have explicitly indicated the sum over the flavor indices and
suppressed the color indices. In the supergravity basis it is customary to use
fields with a different
normalization. The kinetic term for the scalars of the
chiral multiplet is written as
\begin{equation}
\int d^4x ~\sum_{f=1}^{N_F}
\Big\{ K_Q\,D_\mu Q^{\dagger f} \,D^\mu {Q}_f
 + K_{\tilde Q}\, D_\mu {\tilde Q}^f\,D^\mu {\tilde Q}^\dagger_f
\Big\}~,
\label{kinQ}
\end{equation}
where $K_Q$ and $K_{\tilde Q}$ are the K\"ahler metrics.
Upon comparison with
(\ref{kinq}), we see that relation between the fields $q$ and $\tilde q$
appearing in the string vertex operators and the fields
$Q$ and $\tilde Q$ of the supergravity basis is
\begin{equation}
q = \sqrt{K_{Q}^{}}\,Q\quad,\quad
\tilde q = \sqrt{K_{\tilde Q}}\,\tilde Q~.
\label{qQ}
\end{equation}
Actually, the rescalings (\ref{qQ}) apply not only to the scalar components, but
to the entire chiral multiplets.

\section{Instantonic brane effects}
\label{sec:inst_branes}
In this stringy set-up non-perturbative instantonic effects can be
included by adding fractional Euclidean D5 branes (or E5 branes for short)
that completely wrap the internal manifold.
We choose these branes to be identical to the color D$9_a$ branes
in the internal directions (i.e. they transform in the same representation of
the orbifold group; they have, if any, the same magnetization etc.), while they
are point-like in the space-time directions. Thus we call them E$5_a$, and they
provide the stringy representation of ordinary instantons for the gauge theory
on the D$9_a$ branes. Notice, however, that with respect to the gauge
theory living on a different stack of D$9$ branes (like the branes D$9_b$ or
D$9_c$),
the E$5_a$ represent ``exotic'' instantons, whose properties are different from
those of the ordinary gauge theory instantons. Recently, these ``exotic''
configurations  have been investigated
~\cite{Beasley:2005iu}--\cite{Blumenhagen:2007bn}
from various points of view.

Our aim is to use the relation between the non-holomorphic corrections appearing
in the string computation of instantonic effects and the K\"ahler metrics
of the chiral multiplets in the supergravity basis to gain information on the
latter.
To elucidate the physical meaning of these corrections, we will examine
in particular the one-instanton induced ADS/TVY superpotential
\cite{Affleck:1983mk} (see also Ref. \cite{Taylor:1982bp}), present
in the case $N_F = N_a - 1$, whose stringy derivation has been recently
reconsidered in \cite{Akerblom:2006hx,Argurio:2007vq}.
To proceed, let us first review how the instanton contributions to the
superpotential arise in our specific set-up.

\subsection{The instanton moduli}
\label{subsec:inst_moduli}

In presence of the E$5_a$ branes we have new types of
open strings: the E$5_a$/E$5_a$ strings (neutral sector), the
D$9_a$/E$5_a$ or E$5_a$/D$9_a$ strings (charged sector) and the
D$9_b$/E$5_a$ or E$5_a$/D$9_c$ strings (flavored sectors).
The states of such strings do not carry any space-time momentum and
represent moduli rather than dynamical
fields in space-time. The spectrum of moduli is summarized in Table
\ref{tab:moduli}, and the corresponding vertex operators are listed in
Appendix \ref{appsub:vert}. Let us
notice that the states of these strings can carry (discretized) momentum along
the compact directions, when they are untwisted,
{\it i.e.} when they belong to the neutral or charged sectors; such
Kaluza-Klein copies of the moduli represent a genuine string feature.
\begin{table}
\begin{center}
\begin{tabular}{cc|cccc}
\hline\hline
\multicolumn{2}{c}{Sector}  & ADHM  & Meaning & Chan-Paton & Dimension\\
\hline
$\phantom{\vdots}5_a$/$5_a$ & NS & $a_\mu$ & centers & adj. $\mbox{U}(k)$ &
(length)\\
 & &  $D_c$ & Lagrange mult. & $\vdots$ & (length)$^{-2}$\\
 &  R & ${M}^{\alpha}$ &  partners &  $\vdots$ & (length)$^{\frac12}$\\
 &    & $\lambda_{\dot\alpha}$ & Lagrange mult.  & $\vdots$ &
(length)$^{-\frac32}$ \\
\hline
$\phantom{\vdots}9_a/5_a$ & NS &  ${w}_{\dot\alpha}$ & sizes & $N_a \times
\overline{k}$
& (length)\\
$5_a/9_a$ &  & ${\bar w}_{\dot\alpha}$ & $\vdots$ & $k\times \overline{N}_a$ &
$\vdots$\\
$9_a/5_a$ & R & ${\mu}$ & partners & $N_a \times \overline{k}$
& (length)$^{\frac12}$\\
$5_a/9_a$ &  & ${\bar \mu}$ & $\vdots$ &$k\times \overline{N}_a$
&  $\vdots$\\
\hline
$\phantom{\vdots}9_b/5_a$ & R & ${\mu}^\prime$ & flavored & ${N}_F\times
\overline{k}$
&  (length)$^{\frac12}$\\
$5_a/9_c$ &  & ${\tilde \mu}^\prime$ & $\vdots$ & ${k} \times\overline{N}_F
$ & $\vdots$\\
\hline\hline
\end{tabular}
\end{center}
\caption{The spectrum of moduli from the open strings with at least one
boundary attached to the instantonic E$5_a$ branes. See the text for more
details and comments, and Appendix \ref{appsub:vert} for the expressions of the
corresponding emission vertices.}
\label{tab:moduli}
\end{table}

Let us also recall that, in order to yield non-trivial
interactions when $\alpha'\to 0$ \cite{Billo:2002hm}, the emission vertices
of some of the moduli, given in Appendix \ref{appsub:vert}, have to be rescaled
with
factors of  the dimensionful coupling constant on the E$5_a$, namely
$g_{5_a} = g_a/(4\pi^2\alpha')$, with $g_a$ given in (\ref{gym}).
As a consequence, some of the moduli acquire unconventional
scaling dimensions which, however, are the right ones for their interpretation
as parameters of an instanton solution \cite{Dorey:2002ik,Billo:2002hm}.

The neutral moduli which survive the orbifold projection are the four physical
bosonic excitations $a_\mu$ from the NS sector, related to the positions of the
(multi-)centers of the instanton, and three auxiliary excitations $D_c$
($c=1,2,3$). In the R sector, we find two chiral fermionic zero-modes
$M^{\alpha}$, and two anti-chiral ones $\lambda_{\dot\alpha}$. The
${M}^{\alpha}$ are the fermionic partners of the instanton centers. All of these
moduli are $k\times k$ matrices and transform in the adjoint representation of
$\mathrm{U}(k)$. If we write the $k\times k$ matrices ${a}^\mu$ and
${M}^{\alpha}$ as
\begin{equation}
{a}^\mu = x_0^\mu\,\uno_{k\times k} + y^\mu_c\,T^c\quad,\quad
{M}^{\alpha}=\theta^{\alpha}\,\uno_{k \times k } + {\zeta}^{\alpha}_c\,T^c~,
\label{xtheta}
\end{equation}
where $T^c$ are the generators of $\mathrm{SU}(k)$, then the instanton center of
mass, $x_0^\mu$, and its fermionic partners, $\theta^{\alpha }$, can be
identified respectively with the bosonic and fermionic coordinates of the
$\mathcal{N}=1$ superspace.

The charged instantonic sector contains, in the NS sector, two physical bosonic
moduli $w_{\dot\alpha}$ with dimension of (length), related to the size and
orientation in color space of the instanton, and a fermionic modulus $\mu$.
These moduli carry a fundamental $\mathrm{U}(k)$ index and a color one.

\begin{figure}
\begin{center}
\begin{picture}(0,0)%
\includegraphics{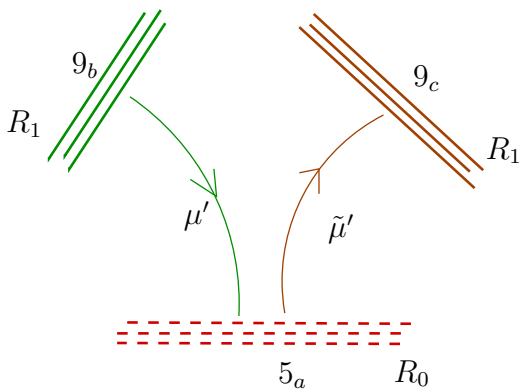}%
\end{picture}%
\setlength{\unitlength}{2171sp}%
\begingroup\makeatletter\ifx\SetFigFont\undefined%
\gdef\SetFigFont#1#2#3#4#5{%
  \reset@font\fontsize{#1}{#2pt}%
  \fontfamily{#3}\fontseries{#4}\fontshape{#5}%
  \selectfont}%
\fi\endgroup%
\begin{picture}(6474,4401)(226,-3784)
\put(3661,-3676){\makebox(0,0)[rb]{\smash{{\SetFigFont{12}{14.4}
{\familydefault}{\mddefault}{\updefault}$5_a$}}}}
\put(5176,-286){\makebox(0,0)[rb]{\smash{{\SetFigFont{12}{14.4}
{\familydefault}{\mddefault}{\updefault}$9_c$}}}}
\put(4201,-2011){\makebox(0,0)[rb]{\smash{{\SetFigFont{12}{14.4}
{\familydefault}{\mddefault}{\updefault}$\tilde{\mu}^\prime$}}}}
\put(2551,-1861){\makebox(0,0)[rb]{\smash{{\SetFigFont{12}{14.4}
{\familydefault}{\mddefault}{\updefault}$\mu^\prime$}}}}
\put(1276,-136){\makebox(0,0)[rb]{\smash{{\SetFigFont{12}{14.4}
{\familydefault}{\mddefault}{\updefault}$9_b$}}}}
\put(4651,-3661){\makebox(0,0)[lb]{\smash{{\SetFigFont{12}{14.4}
{\familydefault}{\mddefault}{\updefault}$R_0$}}}}
\put(5701,-1111){\makebox(0,0)[lb]{\smash{{\SetFigFont{12}{14.4}
{\familydefault}{\mddefault}{\updefault}$R_1$}}}}
\put(226,-811){\makebox(0,0)[lb]{\smash{{\SetFigFont{12}{14.4}
{\familydefault}{\mddefault}{\updefault}$R_1$}}}}
\end{picture}%
\end{center}
\caption{The flavored moduli of instantonic E$5_a$ branes in presence of
D$9_b$ and D$9_c$ branes; see the text for more details.}
\label{fig:5a9b9c}
\end{figure}

In our realization of $\mathcal{N}=1$ SQCD there are two flavored instantonic
sectors
corresponding to the open strings that stretch between the E$5_a$ branes and the
flavor
branes of type $b$ or $c$, depicted in Fig. \ref{fig:5a9b9c}. In both cases the
four non-compact directions have mixed
Neumann-Dirichlet boundary conditions while all the internal complex coordinates
are twisted. As a consequence, there are no bosonic physical zero-modes
in the NS sector and the only physical excitations are fermionic ones
from the R sector. A detailed analysis of the twisted conformal field theory
shows that there are fermionic moduli $\mu'_f$ in the D$9_b$/E$5_a$ strings and
fermionic moduli ${\tilde \mu}'{}^f$ in the E$5_a$/D$9_c$ strings.
They are described respectively by the vertex operators (\ref{vertmup}) and
(\ref{vertmupt}). On the other hand no physical states
survive the GSO projection in the E$5_a$/D$9_b$ and D$9_c$/E$5_a$
sectors. The fermionic moduli $\mu'_f$ and ${\tilde \mu}'{}^f$
are the counterparts of the chiral multiplets $q_f$ and
${\tilde q}^f$ respectively, when the color D$9_a$ branes are
replaced by the instantonic E$5_a$ branes.

The physical moduli we have listed above, collectively denoted by
$\mathcal{M}_k$, are in one-to-one correspondence with the
ADHM moduli of $\mathcal{N}=1$ gauge instantons (for a more detailed discussion
see, for instance, \cite{Dorey:2002ik} and references
therein). In all instantonic sectors
we can construct many other open string states that carry a discretized
momentum along the compact directions and/or have some bosonic or fermionic
string oscillators. These ``massive" states are not physical, {\it
i.e.} they cannot be described by vertex operators of conformal dimension
one; they can, however, circulate in open string
loop diagrams.

\subsection{The instanton induced superpotential}
\label{subsec:inst_sup}
In the sector with instanton number $k$, the effective action for the
gauge/matter fields is obtained by the ``functional'' integral over the
instanton moduli of the exponential of all diagrams with at least part of
their boundary on the E$5_a$ branes, possibly with insertions of
moduli and gauge/matter fields
\cite{Polchinski:1994fq,Green:1997tv,Green:2000ke,Billo:2002hm,
Blumenhagen:2006xt,Billo:2007sw}. In the
semi-classical approximation, only disk diagrams and annuli (the latter with no
insertions) are retained. Focusing on the dependence from the
scalar fields of the chiral multiplets in the Higgs branch, we
have
\begin{equation}
S_{k}=  {\cal C}_k ~\ee^{-\frac{8 \pi^2}{g_a^2}\,k}~
\ee^{\mathcal{A}^\prime_{5_a}} \int d{\mathcal M}_{k}~
\ee^{-\smod}~.
\label{Z1}
\end{equation}
Let us now analyze the various terms in this expression.

${\cal C}_k$ is a normalization factor which
compensates for the dimensions of the integration measure $d{\cal M}_k$, and
may contain numerical constants and powers of the coupling $g_a$.
Its dimensionality is determined by counting the dimensions (measured in units
of $\alpha'$) of the various moduli ${\mathcal M}_{k}$ as given in the previous
subsections, and the result is, up to overall numerical constants,
\begin{equation}
{\mathcal C}_k = \big({\sqrt {\alpha'}}\big)^{-(3N_a-N_F)k}\, (g_a)^{-2N_ak}~.
\label{ck}
\end{equation}
Notice the appearance of the one-loop coefficient $b_1=(3N_a-N_F)$ of the
$\beta$-function of the $\mathcal{N}=1$ SQCD with $N_F$ flavors. The factor of
$(g_a)^{-2N_ak}$ in (\ref{ck}) has been inserted following the discussion of
Ref. \cite{Dorey:2002ik}, but in principle it should also have a stringy
interpretation, probably as a left-over of the cancellation between the bosonic
and fermionic fluctuation determinants when the $\mathcal{N}=1$ gauge action is
normalized as in (\ref{n1}).

As explained in Refs.~\cite{Polchinski:1994fq,Billo:2002hm}, the disk diagrams
with no insertions account for the exponential of (minus) the classical
instanton action $8 \pi^2 k/g_a^2$, where  $g_a$ is interpreted as the
Yang-Mills coupling constant at the string scale. This explains
the second factor in (\ref{Z1}).
The third factor contains $\mathcal{A}^\prime_{5_a}$ which
accounts for the open string annuli diagrams with at least one boundary on the
E$5_a$ branes and no insertions
\cite{Blumenhagen:2006xt,Abel:2006yk,Akerblom:2006hx,Billo:2007sw}. Since the
functional integration over the ADHM moduli ${\cal M}_k$ is explicitly performed
in (\ref{Z1}), to avoid double counting only the contribution of the ``massive''
string excitations has to be taken into account in these annuli: this is the
reason of the $'$ notation which reminds that only the ``massive'' instantonic
string excitations must circulate in the loop.

\begin{figure}
\begin{center}
\begin{picture}(0,0)%
\includegraphics{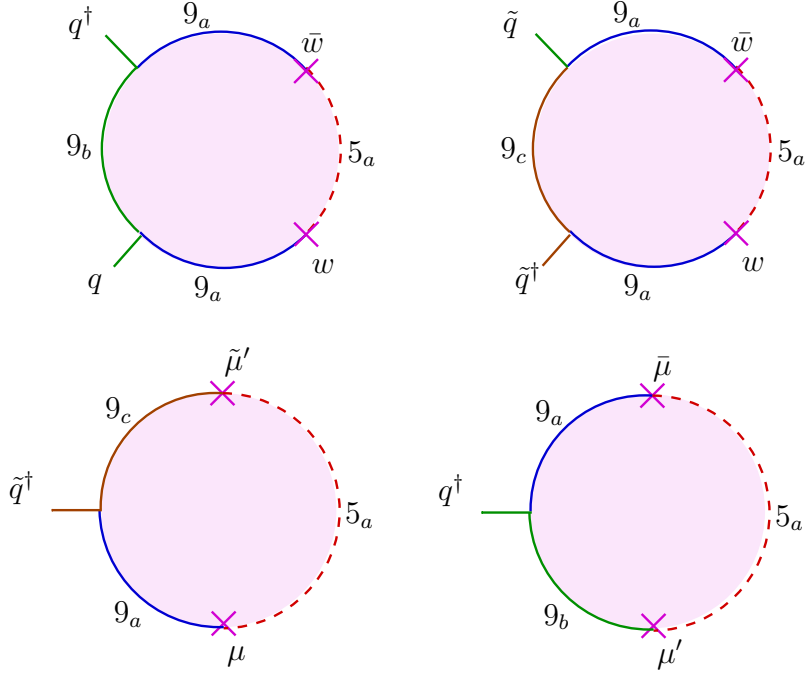}%
\end{picture}%
\setlength{\unitlength}{1973sp}%
\begingroup\makeatletter\ifx\SetFigFont\undefined%
\gdef\SetFigFont#1#2#3#4#5{%
  \reset@font\fontsize{#1}{#2pt}%
  \fontfamily{#3}\fontseries{#4}\fontshape{#5}%
  \selectfont}%
\fi\endgroup%
\begin{picture}(10583,8451)(1,-7849)
\put(1,-5671){\makebox(0,0)[lb]{\smash{{\SetFigFont{12}{14.4}
{\familydefault}{\mddefault}{\updefault}$\tilde{q}^\dagger$}}}}
\put(2746,-7711){\makebox(0,0)[lb]{\smash{{\SetFigFont{12}{14.4}
{\familydefault}{\mddefault}{\updefault}$\mu$}}}}
\put(1321,-7231){\makebox(0,0)[lb]{\smash{{\SetFigFont{12}{14.4}
{\familydefault}{\mddefault}{\updefault}$9_a$}}}}
\put(4231,-5986){\makebox(0,0)[lb]{\smash{{\SetFigFont{12}{14.4}
{\familydefault}{\mddefault}{\updefault}$5_a$}}}}
\put(1216,-4666){\makebox(0,0)[lb]{\smash{{\SetFigFont{12}{14.4}
{\familydefault}{\mddefault}{\updefault}$9_c$}}}}
\put(8146,-7741){\makebox(0,0)[lb]{\smash{{\SetFigFont{12}{14.4}
{\familydefault}{\mddefault}{\updefault}$\mu^\prime$}}}}
\put(6616,-4696){\makebox(0,0)[lb]{\smash{{\SetFigFont{12}{14.4}
{\familydefault}{\mddefault}{\updefault}$9_a$}}}}
\put(6721,-7261){\makebox(0,0)[lb]{\smash{{\SetFigFont{12}{14.4}
{\familydefault}{\mddefault}{\updefault}$9_b$}}}}
\put(9631,-6016){\makebox(0,0)[lb]{\smash{{\SetFigFont{12}{14.4}
{\familydefault}{\mddefault}{\updefault}$5_a$}}}}
\put(5401,-5701){\makebox(0,0)[lb]{\smash{{\SetFigFont{12}{14.4}
{\familydefault}{\mddefault}{\updefault}$q^\dagger$}}}}
\put(4261,-1441){\makebox(0,0)[lb]{\smash{{\SetFigFont{12}{14.4}
{\familydefault}{\mddefault}{\updefault}$5_a$}}}}
\put(2326,-3136){\makebox(0,0)[lb]{\smash{{\SetFigFont{12}{14.4}
{\familydefault}{\mddefault}{\updefault}$9_a$}}}}
\put(2191,299){\makebox(0,0)[lb]{\smash{{\SetFigFont{12}{14.4}
{\familydefault}{\mddefault}{\updefault}$9_a$}}}}
\put(3676,-61){\makebox(0,0)[lb]{\smash{{\SetFigFont{12}{14.4}
{\familydefault}{\mddefault}{\updefault}$\bar w$}}}}
\put(3811,-2836){\makebox(0,0)[lb]{\smash{{\SetFigFont{12}{14.4}
{\familydefault}{\mddefault}{\updefault}$w$}}}}
\put(9661,-1426){\makebox(0,0)[lb]{\smash{{\SetFigFont{12}{14.4}
{\familydefault}{\mddefault}{\updefault}$5_a$}}}}
\put(7726,-3121){\makebox(0,0)[lb]{\smash{{\SetFigFont{12}{14.4}
{\familydefault}{\mddefault}{\updefault}$9_a$}}}}
\put(7591,314){\makebox(0,0)[lb]{\smash{{\SetFigFont{12}{14.4}
{\familydefault}{\mddefault}{\updefault}$9_a$}}}}
\put(9076,-46){\makebox(0,0)[lb]{\smash{{\SetFigFont{12}{14.4}
{\familydefault}{\mddefault}{\updefault}$\bar w$}}}}
\put(9211,-2821){\makebox(0,0)[lb]{\smash{{\SetFigFont{12}{14.4}
{\familydefault}{\mddefault}{\updefault}$w$}}}}
\put(751,179){\makebox(0,0)[lb]{\smash{{\SetFigFont{12}{14.4}
{\familydefault}{\mddefault}{\updefault}$q^\dagger$}}}}
\put(991,-3031){\makebox(0,0)[lb]{\smash{{\SetFigFont{12}{14.4}
{\familydefault}{\mddefault}{\updefault}$q$}}}}
\put(6211,194){\makebox(0,0)[lb]{\smash{{\SetFigFont{12}{14.4}
{\familydefault}{\mddefault}{\updefault}$\tilde{q}$}}}}
\put(6361,-3031){\makebox(0,0)[lb]{\smash{{\SetFigFont{12}{14.4}
{\familydefault}{\mddefault}{\updefault}$\tilde{q}^\dagger$}}}}
\put(721,-1351){\makebox(0,0)[lb]{\smash{{\SetFigFont{12}{14.4}
{\familydefault}{\mddefault}{\updefault}$9_b$}}}}
\put(6181,-1396){\makebox(0,0)[lb]{\smash{{\SetFigFont{12}{14.4}
{\familydefault}{\mddefault}{\updefault}$9_c$}}}}
\put(2716,-4066){\makebox(0,0)[lb]{\smash{{\SetFigFont{12}{14.4}
{\familydefault}{\mddefault}{\updefault}$\tilde{\mu}^\prime$}}}}
\put(8101,-4126){\makebox(0,0)[lb]{\smash{{\SetFigFont{12}{14.4}
{\familydefault}{\mddefault}{\updefault}${\bar \mu}$}}}}
\end{picture}%
\end{center}
\caption{Disk interactions between the moduli and the chiral scalars which
survive in the field theory limit}
\label{fig:1}
\end{figure}

Finally, in the integrand of (\ref{Z1}) we find the moduli action $\smod$. This
can be computed following the procedure explained in Ref. \cite{Billo:2002hm}
from all disk scattering amplitudes involving the ADHM moduli and the scalar
fields $q$ and $\tilde q$ in the limit $\alpha'\to 0$ (with $g_a$ fixed). The
result is
\begin{equation}
\label{smodex}
\begin{aligned}
\smod & = {\rm tr}_k  \Big\{
\ii D_c\Big({\bar
w}_{\dot\alpha}(\tau^c)^{\dot\alpha}_{~\dot\beta}w^{\dot\beta}
+\ii \bar\eta_{\mu\nu}^c \big[{a}^\mu,{a}^\nu\big]\Big)
\\&
- \ii
{\lambda}^{\dot\alpha}\Big(\bar{\mu}{w}_{\dot\alpha}+
\bar{w}_{\dot\alpha}{\mu}  +
\big[a_\mu,{M}^{\alpha}\big]\sigma^\mu_{\alpha\dot\alpha}\Big)\Big\}
\\
& + \tr_k\sum_{f=1}^{N_F} \Big\{ {\bar w}_{\dot\alpha}
\big[q^{\dagger f}{q}_f
 + {\tilde q}^f\tilde{q}^\dagger_f\big]
 w^{\dot\alpha} - \frac{\ii}{2}\, {\bar \mu}\,
q^{\dagger f} \mu'_f +  \frac{\ii}{2}\,
{{\tilde\mu}'}{}^f\tilde{q}^\dagger_f\,
\mu\Big\}~.
\end{aligned}
\end{equation}
The first two lines above have the only effect of implementing the (super) ADHM
constraints, for which $D_c$ and $\lambda^{\dot\alpha}$ act as Lagrange
multipliers.
The last line arises from the disk diagrams which contain insertions of the
chiral scalars and survive in the field theory limit, depicted in Fig.
\ref{fig:1}.

There are other non-zero disk diagrams with moduli and matter fields that
survive in the field theory limit. However, these other diagrams can be related
to the ones in Fig. \ref{fig:1} by means of supersymmetry Ward
identities~\cite{Green:2000ke,Billo:2002hm}. This implies that the complete
result is  obtained simply by replacing in (\ref{smodex}) the
scalars $q$ and $\tilde q$ and their conjugate with the corresponding chiral and
anti-chiral superfields. From now on, we will assume this replacement.
Notice that the multiplets $q$ and $\tilde q$ appear in (\ref{smodex})
differently from their conjugates $q^\dagger$ and ${\tilde q}^\dagger$;
this fact has important consequences on the holomorphicity properties of the
instanton-induced correlators, as we will see later.

In the moduli action (\ref{smodex}), the superspace coordinates $x_0^{\mu}$ and
$\theta^{\alpha}$, defined in (\ref{xtheta}), appear only through superfields
$q(x_0,\theta), \tilde q(x_0,\theta),\ldots$. It is therefore convenient to
separate these coordinates from the remaining centered moduli, denoted by
$\widehat{\mathcal{M}}_k$, and rewrite the effective action (\ref{Z1}) in terms
of
a $k$-instanton induced superpotential $W_k$, namely
\begin{equation}
S_{k}= \int d^4x_0\, d^2\theta ~W_{k}(q,\tilde q)~,
\label{Wk}
\end{equation}
where
\begin{equation}
W_{k}(q,\tilde q)=
{\cal C}_k ~\ee^{-\frac{8 \pi^2}{g_a^2}\,k}~
\ee^{\mathcal{A}^\prime_{5_a}}
\int d{\widehat{\mathcal M}_{k}}~
\ee^{-\smodh}~.
\label{zk3}
\end{equation}
Even if $\smodh$ has an explicit dependence  on $q^\dagger$ and ${\tilde
q}^\dagger$, this dependence disappears upon integrating over $\widehat{\mathcal
M}_{k}$ as a consequence of the cohomology properties of the integration measure
on the instanton moduli space \cite{Hollowood:2002ds,Dorey:2002ik,Billo:2006jm}.
Thus, $W_k(q,\tilde q)$ depends holomorphically on the chiral superfields $q$
and $\tilde q$. However, the annulus amplitude
$\mathcal{A}^\prime_{5_a}$ that appears in the prefactor of Eq. (\ref{zk3})
could introduce a non-holomorphic dependence on the complex and K\"ahler
structure moduli of the compactification space. On the other hand, the
multiplets $q$ and $\tilde q$ have to be rescaled according to Eq. (\ref{qQ}) to
express the result in the supergravity variables, and the
holomorphic Wilsonian renormalization group invariant scale $\Lhol$ has to be introduced.

We will consider the interplay of all these observations in
Section
\ref{sec:relation}, after explicitly evaluating the instantonic annulus
amplitude $\mathcal{A}^\prime_{5_a}$ in Section \ref{subsec:mix_ann}. Before this,
however, we briefly comment on the non-perturbative superpotential for $k=1$.

\subsection{The ADS/TVY superpotential}
\label{subsec:ADS}
The measure $d{\widehat{\mathcal M}_{k}}$ in Eq. (\ref{zk3}) contains many
fermionic zero modes. Among them, the $\lambda^{\dot\alpha}$ are Lagrange
multipliers for the fermionic ADHM constraints but, after enforcing these
constraints, the $\mu$'s, $\bar\mu$'s, $\mu^\prime$'s and ${\tilde\mu}^\prime$'s
must be exactly compensated otherwise the entire integral vanishes. The single
instanton case, $k=1$, is already very interesting. First of all, in this case
it is easy to see that the balancing of the fermionic zero-modes requires that
$N_F = N_a - 1$. After having integrated over the fermions, we are left with a
(constrained) Gaussian integration over the bosonic moduli $w_{\dot\alpha}$ and
$\bar w_{\dot\alpha}$, which can be explicitly performed {\it e.g.} by going to
a region of the moduli space where the chiral fields are diagonal, up to
rows/columns of zeroes. Furthermore, the D-terms in the gauge sector constrain
the superfields to obey $q^{\dagger f}{q}_f={\tilde q}^f\tilde{q}^\dagger_f$, so
that the bosonic integration brings the square of a simple determinant in the
denominator, which cancels the anti-holomorphic contributions produced by the
fermionic integrals. In the end, one finds \cite{Dorey:2002ik} (see also Refs.
\cite{Akerblom:2006hx,Argurio:2007vq})
\begin{equation}
\label{Wk1}
W_{k=1}(q,\tilde q)
= {\cal C}_k ~\ee^{-\frac{8 \pi^2}{g_a^2}\,k}~\ee^{\mathcal{A}^\prime_{5_a}}
\,\frac1{\det\big(\tilde q q\big)}~,
\end{equation}
which has the same form of the ADS/TVY superpotential
\cite{Affleck:1983mk,Taylor:1982bp}.
As we will explicitly see in the following, the prefactor
$\ee^{\mathcal{A}^\prime_{5_a}}$ is crucial to establish the correct holomorphic
properties of this superpotential when everything is expressed in terms of the
supergravity variables (\ref{stu}), the chiral superfields are normalized with
their K\"ahler metrics and the Wilsonian scale $\Lhol$ is introduced.

\section{The mixed annuli}
\label{subsec:mix_ann}
To describe explicitly the instanton induced effects on the low energy effective
action, the only ingredient yet to be specified is the annulus amplitude
$\mathcal{A}_{5_a}$, whose ``primed'' part appears in the equations from Eq.
(\ref{Z1}) on. This amplitude represents the 1-loop vacuum energy of open
strings with at least one end point on the wrapped instantonic branes E$5_a$.
Because of supersymmetry, the annulus amplitude associated to the E$5_a$/E$5_a$
strings identically vanishes, so $\mathcal{A}_{5_a}$ receives contributions only
from mixed annuli with one boundary on the E$5_a$'s and the other on the D$9$
branes. In particular, the 1-loop contribution of the charged instantonic open
strings is denoted as
\begin{equation}
\label{a5a2or}
\mathcal{A}_{5_a;9_a} = \mathcal{A}(9a/5a) + \mathcal{A}(5a/9a)~,
\end{equation}
where on the r.h.s. we distinguish the contributions of the D$9_a$/E$5_a$ and
E$5_a$/D$9_a$. Similarly, for the flavored instantonic open strings
\begin{equation}
\label{bc5a2or}
\mathcal{A}_{5_a;9_b} = \mathcal{A}(9b/5a) + \mathcal{A}(5a/9b)
\quad\mbox{and}\quad
\mathcal{A}_{5_a;9_c} = \mathcal{A}(9c/5a) + \mathcal{A}(5a/9c)
\end{equation}
for the two different stacks of flavor branes used to engineer $\mathcal{N}=1$
SQCD.

It has been noticed in the literature \cite{Abel:2006yk,Akerblom:2006hx} that
the computation of mixed annuli is related to the stringy computation of the
1-loop threshold corrections to the coupling of the color gauge group living on
the D$9_a$. In Ref. \cite{Billo:2007sw} we showed that this relation is
explained by the fact that, in a supersymmetric theory, the mixed annuli compute
just the running coupling by expanding around the classical instanton
background, namely
\begin{equation}
\label{a5tog}
\mathcal{A}_{5_a} = -\frac{8\pi^2 k}{g_a^2(\mu)}\,\Bigg|_{\mathrm{\,at\,1-loop}}~.
\end{equation}
Applying this argument to our system and keeping distinct the charged and
flavored sectors, we expect therefore to find
\begin{subequations}
\label{ma_running}
\begin{align}
\mathcal{A}_{5_a;9_a} & =
- 8\pi^2k\left(\frac{3N_a}{16 \pi^2} \log(\alpha'\mu^2)
+ \Decol\right)~,
\label{macolor}\\
\mathcal{A}_{5_a;9_b}+\mathcal{A}_{5_a;9_c} & =
- 8\pi^2k\left(-\frac{N_F}{16 \pi^2} \log(\alpha'\mu^2)
+ \Deflav\right)~.
\label{maflavor}
\end{align}
\end{subequations}
In these expressions, $\mu$ is the scale that regularizes the IR divergences of
the annuli amplitudes%
\footnote{These open string annulus amplitudes exhibit both UV and IR
divergences. The UV divergences, corresponding to IR divergences in the dual
closed string channel, cancel in consistent tadpole-free models; even if in this
paper we take only a local point of view, we assume that globally the closed
string tadpoles are absent so that we can ignore the UV divergences.} due to the
massless states circulating in the loop, and the coefficients of the logarithms
arise by counting (with appropriate sign and weight) the bosonic and fermionic
ground states of mixed open strings with one end point on the E$5_a$s branes,
{\it i.e.} the charged and flavored instanton moduli that we listed in
Section
\ref{sec:inst_branes}. This counting agrees, as it should, with the 1-loop
$\beta$-function coefficients that are appropriate, respectively, for the gauge
and the flavor multiplets. Let us now describe the explicit form of the various
annuli amplitudes.

\paragraph{Charged sector}
For a given open string orientation, we have
\begin{equation}
\label{ma0}
\mathcal{A}(9_a/5_a) = \int _0^\infty \frac{d\tau}{2\tau}\left[
\Tr_{\mathrm{NS}}\left(P_{\mathrm{GSO}}^{(9_a/5_a)} \,P_{\mathrm{orb.}}\,
q^{L_0}\right)
- \Tr_{\mathrm{R}}\left(P_{\mathrm{GSO}}^{(9_a/5_a)}\,P_{\mathrm{orb.}}\,
q^{L_0}\right)\right]~,
\end{equation}
where $q= \exp(-2\pi\tau)$, $P_{\mathrm{GSO}}^{(9_a,5_a)}$ is the appropriate
GSO projector, and
\begin{equation}
\label{porb}
P_{\mathrm{orb.}}= \frac 14\left(1 + \sum_{i=1}^3 h_i\right)
\end{equation}
is the orbifold projector, with $h_i$ being the three non-trivial elements of
the
$\mathbb{Z}_2\times\mathbb{Z}_2$ orbifold action of our
background. Each element $h_i$ is in fact the generator of a $\mathbb{Z}_2$
subgroup which leaves invariant the $i$-th torus (see Appendix
\ref{appsub:geom}). The
corresponding term in the
amplitude is therefore identical in form to the one encountered in the
computation of the $9a/5a$ amplitude in a $\mathcal{N}=2$ background
$\mathcal{T}_2^{(j)}\times \mathcal{T}_2^{(k)}$ (with $j,k\not=i$). This
computation is described, for instance, in Section 4 of Ref. \cite{Billo:2007sw}
to
which we
refer for notations and details. It turns out that the GSO projection in the R
sector has to be defined differently for the two string orientations
(see Appendix \ref{appsub:vert})
so that the amplitude $\mathcal{A}(9a/5a)$ vanishes, and one remains with
 \begin{equation}
\label{atotfin}
\mathcal{A}_{5_a;9_a} = \mathcal{A}(5a/9a) = N_a k
\sum_{i=1}^3 \,\int_0^\infty
\frac{d\tau}{2\tau}
\, \mathcal{Y}^{(i)}~.
\end{equation}
In the end all string excitations cancel and only the zero-modes contribute:
they correspond to the charged instanton moduli listed in Table
\ref{tab:moduli}, and their Kaluza-Klein partners on the torus
$\mathcal{T}_2^{(i)}$ fixed by the element $h_i$ of the orbifold group; these
states reconstruct the sum
\begin{equation}
\label{ma3}
\mathcal{Y}^{(i)}\equiv
\sum_{(r_1 , r_2) \in \mathbb{Z}^2} q^{
\frac{ | r_{1} U^{(i)}-r_{2} |^2}{U_{2}^{(i)}T_{2}^{(i)}
}}~.
\end{equation}
The integration over the modular parameter can be done
\cite{Lust:2003ky,Billo:2007sw} with the assumption that the UV divergence for
$\tau\to 0$, which corresponds to an IR divergences in the closed string
channel, cancel in a globally consistent, {\it i.e.} tadpole-free, model (of
which here we are considering just the ``local'' aspects on some given stacks of
branes far from the orientifold planes). The IR divergence for $\tau\to\infty$
requires the introduction of cut-offs $m_{(i)}$ which are conveniently taken to
be complex, as advocated in \cite{Di Vecchia:2003ae,Di
Vecchia:2005vm,Billo:2007sw}. The resulting amplitude is then
\begin{equation}
 \label{A5finaa1}
\mathcal{A}_{5_a;9_a} = - \,\frac{N_ak}{2} \,
\sum_{i=1}^3 \,
\left( \log(\alpha^\prime m_{(i)}^2) +
\log \big(U_2^{(i)}T_2^{(i)} |\eta(U^{(i)}|^4 \big)\right)~.
\end{equation}
Choosing \cite{Billo:2007sw}
\begin{equation}
m_{(i)}\,=\, \mu\,\ee^{\ii\varphi_{(i)}}\,=\, \mu\,\ee^{2\ii\, {\rm
arg}(\eta(U^{(i)}))}~,
\label{ircutoff}
\end{equation}
the final result is
\begin{equation}
 \label{A5finaa}
\mathcal{A}_{5_a;9_a} = - 8 \pi^2 k
\left[\frac{3N_a}{16\pi^2} \log(\alpha^\prime \mu^2) + \frac{N_a}{16\pi^2}
\sum_{i=1}^3
\log \Big(U_2^{(i)}T_2^{(i)} (\eta(U^{(i)})^4\Big)\right]~,
\end{equation}
which is of the expected form (\ref{macolor}).

\paragraph{Flavored sectors}
The amplitude $\mathcal{A}_{5_a;9_b}$ receives contributions from the two
possible orientations of the open strings, as described in Eq. (\ref{bc5a2or}).
These contributions, and therefore also the total amplitude
$\mathcal{A}_{5_a;9_b}$, are defined in perfect analogy with Eq. (\ref{ma0}), in
particular they contain the orbifold projector $P_{\rm orb}$ given in
(\ref{porb}). Taking into account that the D$5_a$ branes transform according to
the trivial representation of the orbifold group, while the D$9_b$ and D$9_c$
transform according to the representation $R_1$ defined in Table \ref{frac3}, we
can make explicit the orbifold action on the Chan-Paton factors. We can then
write the total amplitude as the sum of four sectors corresponding to the
insertions of the various group elements as follows%
\footnote{In this notation the identity element
$e$ corresponds to $h_0$.}:
\begin{equation}
{\cal{A}}_{5_a;9_b}
\equiv \frac 14 \sum_{I=0}^3  R_1(h_I)\, {\cal{A}}^{h_I}_{5_a;9_b}
= \frac 14 \left\{{\cal{A}}^{e}_{5_a;9_b} +
{\cal A}_{5_a;9_b}^{h_1} -{\cal A}_{5_a;9_b}^{h_2} - {\cal A}_{5_a;9_b}^{h_3}
\right\}~.
\label{somma34}
\end{equation}
The annulus amplitudes ${\cal{A}}^{h_I}_{5_a;9_b}$ take into account the action
of the orbifold elements $h_I$ on the string fields $Z^i$ and $\Psi^i$, in the
various sectors, as described in Appendix \ref{appsub:geom}. Such amplitudes are
computed in detail in Appendix \ref{appsub:annuli}; here we simply write the
final results. In the untwisted sector we find
\begin{equation}
\label{z59bis}
{\cal{A}}_{5_a;9_b}^{e} = \frac{\ii\, k\,{N_F}}{2 \pi}
\int_{0}^{\infty} \frac{d\tau}{2 \tau}
\left[\,\sum_{i=1}^3 R_1(h_i)\,\partial_{z}
\log \theta_1 (z| \ii \tau )\big|_{z= \ii \tau \nu^{(i)}}
\right]~,
\end{equation}
while in the three twisted sectors we have
\begin{equation}
\label{59tw}
\begin{aligned}
{\cal A}_{5_a;9_b}^{h_i} = & \frac{\ii\, k\,{N_F}}{2 \pi}
\int_{0}^{\infty} \frac{d\tau}{2 \tau}\\
& \times \Bigg[
\partial_{z}\log\theta_1(z|\ii\tau)\big|_{z=\ii\tau\nu^{(i)}}
+ R_1(h_{i})
\sum_{j\neq i=1}^3 R_1(h_j) \,\partial_{z}\log
\theta_2(z|\ii\tau)\big|_{z= \ii\tau\nu^{(j)}}\Bigg]~.
\end{aligned}
\end{equation}
It is worth pointing out that the contribution of the odd spin-structure, which
is divergent due to the superghost zero-modes, actually cancels out when in each
sector we sum over the two open string orientations, leading to finite and
well-defined expressions.

Inserting the amplitudes (\ref{z59bis}) and (\ref{59tw}) in (\ref{somma34}), we
find
\begin{equation}
{\cal{A}}_{5_a;9_b}=\frac{\ii\, k\,{N_F}}{4 \pi}
\int_{0}^{\infty} \frac{d\tau}{2 \tau}
\left\{\,\sum_{i=1}^3 R_1(h_i)\,\partial_{z}
\log \big[\theta_1 (z| \ii \tau )\,\theta_2 (z| \ii \tau )\big]\big|_{z= \ii
\tau \nu^{(i)}}
\right\}~.
\label{d5d9fin1}
\end{equation}
Then, if we use the identity
\begin{eqnarray}
\theta_1 (z| i \tau )\, \theta_2 (z| i \tau ) =
\theta_1 (2z| 2i \tau )\,\prod_{n=1}^{\infty} \left(\frac{ 1 - q^{2n}}{1 +
q^{2n}} \right)~,
\label{1+2=1}
\end{eqnarray}
where $q = \exp(-2\pi\tau)$, it is easy to see that the total flavored
amplitude (\ref{d5d9fin1}) becomes
\begin{equation}
{\cal A}_{5_a;9_b} = \frac{\ii\, k\,{N_F}}{2 \pi}
\int_{0}^{\infty} \frac{d\tau}{2 \tau}
\left[\,\sum_{i=1}^3 R_1(h_i)\,\partial_{z}
\log \theta_1 (z| \ii \tau )\big|_{z= \ii \tau \nu^{(i)}}
\right]~.
\label{d5d9fin2}
\end{equation}
Notice that this is identical to the contribution (\ref{z59bis}) of the
untwisted sector. This means that the flavored amplitude of the orbifold theory
is the same as the one without the orbifold, so that the $\mathcal{N}=1$
structure realized with the magnetic fluxes is fully preserved by the orbifold
projection. The mixed amplitude (\ref{d5d9fin2}) agrees with the quadratic term
in the gauge field $f$ of the annulus amplitude ${\cal A}_{9a;9b}(f)$ computed
in Ref.~\cite{Lust:2003ky} to evaluate the gauge threshold corrections in
intersecting brane models (see also Refs.
\cite{Abel:2006yk,Akerblom:2006hx,Akerblom:2007np}).

We now need to evaluate the integral over $\tau$ that appears in Eq.
(\ref{d5d9fin2}). It is not difficult to realize that this integral is divergent
both in the ultraviolet ($\tau\to 0$) and in the infrared ($\tau\to\infty$). The
ultraviolet divergence can be eliminated by considering tadpole free models as
mentioned above, while the infrared divergence can be cured by introducing, for
example, a regulator $R(\tau)= \big(1-\ee^{- 1/ (\alpha'  m^2 \tau) }\big)$ with
the cut-off $m\rightarrow 0$. The original evaluation \cite{Lust:2003ky} of the
$\tau$ integral appearing in (\ref{d5d9fin2}) has been recently revisited in
Ref. \cite{Akerblom:2007np}. Using this revised result in our case%
\footnote{See
in particular Eq. (3.16) of Ref. \cite{Akerblom:2007np} with all numerical
additive constants absorbed in a redefinition of the cut-off $m$.}, we obtain
\begin{equation}
 \label{A5finab}
\mathcal{A}_{5_a;9_b} = 8\pi^2 k
\left(\frac{N_F}{32\pi^2} \log(\alpha^\prime
m^2) +   \frac{N_F}{32\pi^2} \log \Gammab_{ba}\right)~,
\end{equation}
where
\begin{equation}
 \label{Gammab}
\Gammab_{ba} = \frac{\Gamma(1-\nu_{ba}^{(1)})}{\Gamma(\nu_{ba}^{(1)})}
\frac{\Gamma(\nu_{ba}^{(2)})}{\Gamma(1 - \nu_{ba}^{(2)})}
\frac{\Gamma(\nu_{ba}^{(3)})}{ \Gamma(1 - \nu_{ba}^{(3)})}~.
\end{equation}
Considering also the contribution of the flavor branes of type $c$ that are
characterized by twist angles $\nu_{ac}^{(i)}$, and writing
$m=\mu\,\ee^{\ii\varphi}$, for our realization of ${\cal{N}}=1$ SQCD the total
flavored annulus amplitude is
\begin{equation}
 \label{A5finac}
\mathcal{A}_{5_a;9_b} +\mathcal{A}_{5_a;9_c} = 8\pi^2 k
\left(\frac{N_F}{16\pi^2} \big[\log(\alpha^\prime
\mu^2) +  2 \ii   \varphi
\big] + \frac{N_F}{32\pi^2} \log
\left( \Gammab_{ba} \, \Gammab_{ac} \right) \right)~,
\end{equation}
which is indeed of the expected form (\ref{maflavor}).

\section{Relation to the matter K\"ahler metric}
\label{sec:relation}
In this section we elaborate on the previous results. In particular we will
rewrite the annulus amplitudes (\ref{A5finaa}) and (\ref{A5finac}) in terms of
the variables (\ref{stu}) of the supergravity basis in order to obtain information
on the K\"ahler metrics for the fundamental $\mathcal{N}=1$ chiral multiplets, and then
check that the instanton induced superpotential $W_k$ acquires the correct
holomorphy properties required by $\mathcal{N}=1$ supersymmetry.

\subsection{Holomorphic coupling redefinition}
\label{subsec:hol_red}
As remarked already in Refs.~\cite{Kaplunovsky:1994fg,Louis:1996ya},
the UV cutoff that has to be used in the field theory analysis
of a string model is the four-dimensional Planck mass $M_P$, which
is related to $\alpha'$ as follows:
\begin{equation}
\label{mp}
M_P^2\,=\,\frac{1}{\alpha'}\,{\rm e}^{-\phi_{10}}\,s_2~,
\end{equation}
where $\phi_{10}$ is the ten-dimensional dilaton. In terms of this cut-off, Eqs.
(\ref{A5finaa}) and (\ref{A5finac}) become, respectively,
\begin{equation}
\label{A5finaa2}
\mathcal{A}_{5_a;9_a} = - 8 \pi^2 k
\left(\frac{3N_a}{16\pi^2} \log\frac{\mu^2}{M_P^2}\,
+ \Decolt\right)
\end{equation}
with
\begin{equation}
\label{delc}
 \Decolt\,=\,
\frac{N_a}{16\pi^2}  \left( 3 \log ({\rm e}^{-\phi_{10}}\,s_2)\,+\,
\sum_{i=1}^3 \log \left(U_2^{(i)} T_2^{(i)}
(\eta(U^{(i)})^4\right)\right)~,
\end{equation}
and
\begin{equation}
\label{A5finab2}
\mathcal{A}_{5_a;9_b} + \mathcal{A}_{5_a;9_c}
\,=\,-\,8\pi^2k\left(-\frac{N_{F}}{16 \pi^2} \log \frac{\mu^2}{M_P^2}\,+\,
\Deflavt\right)
\end{equation}
with
\begin{equation}
\label{del}
\Deflavt \,=\,-\,
\frac{N_{F}}{16 \pi^2} \left( \log ({\rm e}^{-\phi_{10}}\,s_2)\,+\,2\ii\varphi
\,+\,
\frac{1}{2} \log \big( \Gammab_{ba}\,\Gammab_{ac}  \big) \right)~.
\end{equation}
Since in $\Deflavt$ there are no analytic terms, we can consistently set
$\varphi=0$ in the following.

We now rewrite the above expressions in terms of the geometrical variables of
the supergravity basis. For the charged amplitude $\mathcal{A}_{5_a;9_a}$ the
procedure is very similar to the one we have applied in the ${\mathcal N}=2$
case \cite{Billo:2007sw}. In fact, using the tree-level relation between the string and
the supergravity moduli given in (\ref{stu}) and the bulk K\"ahler potential
(\ref{kpot}), Eq. (\ref{A5finaa2}) can be recast in the following form:
\begin{equation}
\label{A5aa4}
\mathcal{A}_{5_a;9_a}
\,=k \left[- \frac{3N_{a}}{2} \log \frac{\mu^2}{M_P^2}
- N_a \sum_{i=1}^3 \log \left(\eta(u^{(i)})^2\right)+\frac{N_a}2 \,K
+N_a \log g_a^2
\right]~.
\end{equation}
Turning to the flavored amplitude (\ref{A5finab2}), we easily see that
it can be rewritten as follows:
\begin{equation}
{\cal A}_{5_a;9_b}+{\cal A}_{5_a;9_c} = k\left[\frac{N_{F}}{2} \log
\frac{\mu^2}{M_{P}^{2}}
- \frac{N_{F}}{2} K  + \frac{N_{F}}{2} \log (\mathcal{Z}_Q \mathcal{Z}_{\widetilde
Q})\right]~,
\label{kalo}
\end{equation}
where the quantities $\mathcal{Z}_Q$ and $\mathcal{Z}_{\widetilde Q}$, defined through
the equation
\begin{equation}
\label{gkq}
\log ({\rm e}^{-\phi_{10}}\,s_2)\,+\, \frac{1}{2} \log
\left( \Gammab_{ba} \Gammab_{ca} \right) \,=\,
- K +  \log (\mathcal{Z}_Q \mathcal{Z}_{\widetilde Q})~,
\end{equation}
are explicitly given by
\begin{equation}
\mathcal{Z}_Q =  \big(4\pi\,s_{2}\big)^{-\frac14}\,
\big( t_{2}^{(1)} t_{2}^{(2)}  t_{2}^{(3)} \big)^{-\frac14}\,
\big( u_{2}^{(1)}   u_{2}^{(2)} u_{2}^{(3)}\big)^{-\frac12}
\,\big(\Gammab_{ba}\big)^{\frac12}~,
\label{KQ}
\end{equation}
and
\begin{equation}
{\mathcal{Z}}_{\widetilde{Q}} =  \big(4\pi\,s_{2}\big)^{-\frac14}\,
\big( t_{2}^{(1)} t_{2}^{(2)}  t_{2}^{(3)} \big)^{-\frac14}\,
\big( u_{2}^{(1)}   u_{2}^{(2)} u_{2}^{(3)}\big)^{-\frac12}
\,\big(\Gammab_{ac}\big)^{\frac12}~.
\label{KQtilde}
\end{equation}
Eqs. (\ref{A5aa4}) and (\ref{kalo}) can be combined in the general formula at 1-loop
\begin{equation}
\mathcal{A}
=k\left[ -\frac{b}2\,\log \frac{\mu^2}{M_P^2}\,+\,f^{(1)}\,+\,\frac{c}{2}\,K
\,-\,T(G_a)\,\log\left(\frac{1}{g_a^2}\right)+\sum_r n_r\,T(r)\,\log \mathcal{Z}_r \right]
\label{kl}
\end{equation}
where $f^{(1)}$ is a holomorphic function and
\begin{equation}
\label{bc}
\begin{aligned}
&T(r)\,\delta_{AB}=\Tr_r\big(T_AT_B\big)\quad,\quad
T(G_a)=T(\mathrm{adj})~,
\\
&b=3\,T(G_a)-\sum_r n_r\, T(r)\quad,\quad c=T(G_a)-\sum_r n_r\,T(r)
\end{aligned}
\end{equation}
with $T_A$ being the generators of the gauge group $G_a$ and $n_r$ the number of
$\mathcal{N}=1$ chiral multiplets transforming in the representation
$r$. Indeed, Eq. (\ref{A5aa4}) is obtained from Eq. (\ref{kl}) when we consider the $\mathcal
N=1$ vector multiplet ({\it i.e.} $b= 3 N_a$ and $c= N_a$), and
take
\begin{equation}
\label{ff}
f^{(1)} =-  \, N_a \sum_{i=1}^3 \log \left(\eta(u^{(i)})^2\right)~.
\end{equation}
On the contrary Eq. (\ref{kalo}) is  obtained from Eq. (\ref{kl}) by
considering the fundamental matter fields of $\mathcal{N}=1$ SQCD with $N_F$
flavors ({\it i.e.} $b = c= -N_F$), taking $f^{(1)}=0$ and identifying
$\mathcal{Z}_r$ with $\mathcal{Z}_Q$ and
${\mathcal{Z}}_{\widetilde{Q}}$ of Eqs. (\ref{KQ}) and
(\ref{KQtilde}).

In view of the relation (\ref{a5tog}), we see that by
adding the disk contribution to the above annulus
amplitude one obtains the running coupling constant of the effective
theory in the 1-loop approximation, namely
\begin{equation}
\mathcal{A}_{\mathrm{1-loop}} = -\frac{8\pi^2k}{g_a^2} + \mathcal{A}
\label{new1loop}
\end{equation}
with $\mathcal{A}$ given by (\ref{kl}), {\it i.e.} by the sum of Eqs. (\ref{A5aa4})
and (\ref{kalo}).
On the other hand, according to Refs.
\cite{Dixon:1990pc,Kaplunovsky:1994fg,Louis:1996ya,Shifman:1986zi}
this has to be expressed in terms of the Wilsonian gauge coupling
${\tilde g}_a$, the (tree-level) bulk K\"ahler potential $K$ and
the (tree-level) K\"ahler metrics $K_r$ of the chiral multiplets
in the representation $r$ of the gauge group $G_a$ as follows
\begin{equation}
\mathcal{A}_{\mathrm{1-loop}} = -\frac{8\pi^2k}{{\tilde g}_a^2}
+k\left[ -\frac{b}2\,\log \frac{\mu^2}{M_P^2}+f^{(1)}+\frac{c}{2}\,K
-T(G_a)\log\left(\frac{1}{{\tilde g}_a^2}\right)
+\sum_r n_r\,T(r)\log K_r \right]
\label{kl1}
\end{equation}
with $f^{(1)}$ being a holomorphic function, and $b$ and $c$ defined as in (\ref{bc}).

Notice that the gauge coupling constant $g_a$ obtained from the
disk amplitude may not coincide with the Wilsonian coupling ${\tilde g}_a$
appearing in (\ref{kl1}): in general there may be loop effects,
related to sigma-model anomalies in the low energy
supergravity theory \cite{Derendinger:1991hq}, so that
\begin{equation}
\frac{1}{g_a^2} = \frac{1}{{\tilde g}_a^2}+\frac{\delta}{8\pi^2}~.
\label{deltags}
\end{equation}
Since ${\tilde g}_a$ is the Wilsonian coupling, it has to be (the imaginary part of) a chiral
field: at tree-level this is indeed what happens (see Eq.
(\ref{gym})), but such a relation may be spoiled by loop corrections leading to $\delta\not=0$.
Furthermore, ${\tilde g}_a$ runs only at 1-loop and its
$\beta$-function is given by
\begin{equation}
\beta_{\mathrm{W}}({\tilde g}_a) = -\frac{b_1\,{\tilde g}_a^3}{16\pi^2}~.
\label{betaw}
\end{equation}
Comparing Eq. (\ref{kl1}) with the string expression (\ref{new1loop}) for
$\mathcal{A}_{\mathrm{1-loop}}$, we see:
\begin{itemize}
\item that in the right hand side of
Eq. (\ref{kl}) the $\log(1/g_a^2)$ term can be replaced by $\log(1/{\tilde g}_a^2)$
since the difference yields a higher order correction;
\item
that $f^{(1)}$ is given by (\ref{ff}) and
\item
that if $\delta$ contains a term $\delta^{(0)}$ of order ${\tilde{g}_a}^{\,0}$,
the tree-level K\"ahler metrics
of the chiral fields are given by
\begin{equation}
K_r = \mathcal{Z}_r \mathcal{X}_r
\label{ZX}
\end{equation}
where the non-holomorphic factors $\mathcal{X}_r$ are such that
\begin{equation}
\delta^{(0)} + \sum_r n_r\,T(r)\log \mathcal{X}_r=0~.
\label{deltaX}
\end{equation}
\end{itemize}
Note that if $\delta$ is of higher order in $\tilde{g}_a$, {\it i.e.} if $\delta^{(0)}=0$,
then $\mathcal{X}_r=1$ and the tree-level K\"ahler metric
of the chiral multiplets reduces to $\mathcal{Z}_r$ (see Eqs. (\ref{KQ}) and (\ref{KQtilde})),
that is to what can be directly read from the string annulus amplitude (\ref{kl}).

In the following subsection we will check the consistency of
this result by showing that
the instanton induced superpotential has the correct holomorphy properties
required by $\mathcal{N}=1$ supersymmetry when everything is expressed in the
appropriate variables of the low energy effective action.
Moreover in Section \ref{sec:yukawa} we will compare our findings against the
holomorphy properties of the Yukawa superpotential computed in Ref.
\cite{Cremades:2004wa} for systems of magnetized D9 branes in the field theory
limit.

\subsection{Field redefinitions and the instanton induced superpotential}
\label{subsec:fr_ADS}

The threshold corrections $\Decolt$ and $\Deflavt$, and especially their
non-holomorphic parts, play an important r\^ole since they are related to the
``primed'' part of the annulus amplitude that appears in the prefactor of the
instantonic correlators; in fact
\begin{equation}
\mathcal{A}^\prime_{5_a} = -  8 \pi^2 k
\left(\Decolt + \Deflavt \right)~.
\label{aprime}
\end{equation}
In Ref. \cite{Akerblom:2007uc} it has been suggested that some of the terms of
$\mathcal{A}^\prime_{5_a}$ are related to the rescalings of the fields appearing
in the instanton induced correlator that are necessary in order to have a pure
holomorphic expression. In Ref. \cite{Billo:2007sw} we have showed in detail
that for $\mathcal{N}=2$ models, where the instantons determine corrections to
the gauge prepotential, this is indeed what happens. Here we show that the same
is true also for the instanton-induced superpotential for $\mathcal{N}=1$
theories, thus clarifying the general procedure.

We concentrate in the one-instanton case ($k=1$), where one finds, for $N_F =
N_a-1$, the instanton-induced ADS/TVY-like superpotential of \eq{Wk1}.
For $k=1$, the ``primed'' amplitude (\ref{aprime}) explicitly reads
\begin{equation}
\label{a5p}
\mathcal{A}_{5_a}^\prime =
- N_a \sum_{i=1}^3 \log \left(\eta(u^{(i)})^2 \right) + N_a \log g_a^2
+ \frac{N_a - N_F}{2} K + \frac{N_F}{2} \log(\mathcal{Z}_Q \mathcal{Z}_{\widetilde Q})~.
\end{equation}
Using this expression in Eq. (\ref{Wk1}) and introducing
the K\"ahler metrics $K_Q=\mathcal{Z}_Q \mathcal{X}_Q$
and $K_{\widetilde Q}=\mathcal{Z}_{\widetilde Q}\mathcal{X}_{\widetilde
Q}$, with $\mathcal{Z}_{Q}$ and $\mathcal{Z}_{\widetilde Q}$ given in
(\ref{KQ}) and (\ref{KQtilde}), and $\mathcal{X}_{Q}$ and $\mathcal{X}_{\widetilde Q}$
such that
\begin{equation}
\delta^{(0)} + \frac{N_F}{2}\,\log(\mathcal{X}_Q \mathcal{X}_{\widetilde
Q})=0~,
\end{equation}
we obtain
\begin{equation}
W_{k=1} = \ee^{{K}/2}\,\prod_{i=1}^3\left(\eta(u^{(i)})^{-2 N_a}\right)\,
\Big((\sqrt{\alpha'})^{-(2 N_a +1)}\,\ee^{-\frac{8 \pi^2}{{\widetilde
g}_a^2}}\Big)\,
\big(K_Q K_{\tilde
Q}\big)^{\frac{N_a-1}2}\,\frac{1}{\det(\tilde q q)}~.
\label{Wk12}
\end{equation}
where we have used Eq. (\ref{deltags}) and the fact that
$N_F=N_a-1$.
If we now introduce the chiral multiplets $Q$ and $\tilde Q$ in
the supergravity basis through the rescalings (\ref{qQ}),
and the holomorphic renormalization group
invariant scale through the $\beta$-function (\ref{betaw}), namely
\begin{equation}
\Lhol^{b} = (\sqrt{\alpha'})^{-b}\,\ee^{-\frac{8 \pi^2}{{\widetilde
g}_a^2}}~,
\end{equation}
we find
\begin{equation}
\label{Wk1hol}
W_{k=1} =
\ee^{K/2}\,\prod_{i=1}^3\left(\eta(u^{(i)})^{-2 N_a}\right)\,
\Lhol^{2 N_a +1}\,\frac{1}{\det(\widetilde Q \,Q)} \equiv \ee^{K/2} \,
\Lholt^{\,2 N_a +1}\,\frac{1}{\det(\widetilde Q Q)}
~.
\end{equation}
In the last step we have absorbed the moduli dependent factors of
$\eta(u^{(i)})$ with a holomorphic redefinition of the Wilsonian scale
$\Lhol$ into $\Lholt$.

The final form of \eq{Wk1hol} is the correct one for a holomorphic
ADS/TVY superpotential term in a non-trivial background. The factor
of $\ee^{K/2}$ is the contribution of the bulk K\"ahler potential,
while the remaining part
\begin{equation}
{\widehat W}_{k=1} = \Lholt^{\,2 N_a +1}\,\frac{1}{\det(\widetilde Q Q)}
\label{hatW12}
\end{equation}
is a holomorphic expression in the appropriate variables of the Wilsonian
scheme. Thus, the various pieces of the ``primed'' instantonic
annulus amplitude $\mathcal{A}_{5_a}^\prime$ have conspired to reproduce the
required factors to obtain a fully holomorphic ADS/TVY superpotential
${\widehat W}_{k=1}$.

\section{Comparison with the Yukawa couplings}
\label{sec:yukawa}
It is well known that the K\"ahler metrics of the chiral multiplets play a key
r\^ole in relating the holomorphic superpotential couplings in the effective
supergravity Lagrangian to the physical Yukawa couplings of the canonically
normalized matter fields. This relation represents therefore a possible test
on the structure of the K\"ahler metrics $K_Q$ and $K_{\tilde Q}$.

Let us recall some basic points, and set up appropriate notations. When various
stacks of branes, of types $a,b,c,\ldots$, are present, there are chiral
multiplets arising from the massless open strings stretching between them. We
will denote as $q^{ba}$ the chiral multiplet (as well as the scalar therein)
coming from the D$9_b$/D$9_a$ strings, which we formerly indicated as $q$, and
as $q^{ac}$ the chiral multiplet corresponding to  D$9_a$/D$9_c$ strings, which
was previously indicated as $\tilde q$. We will then similarly have the
multiplets $q^{cb}$, \ldots . The corresponding multiplets in the
``supergravity'' basis will be denoted as $Q^{ba}$, $Q^{ac}$, $Q^{cb}$, \ldots,
and their K\"ahler metrics will be $K_{ba}$ (formerly $K_Q$), $K_{ac}$ (formerly
$K_{\tilde Q}$) and so on. These metrics will contain the appropriate factors
$\Gammab_{ba}$, $\Gammab_{ac}$, $\Gammab_{cb}$, \ldots~, which in
turn are given by the analogue of \eq{Gammab} in terms of the twist angles
$\nu^{(i)}_{ba}$, $\nu^{(i)}_{ac}$, $\nu^{(i)}_{cb}$,
\ldots~.

\begin{figure}
\begin{center}
\begin{picture}(0,0)%
\includegraphics{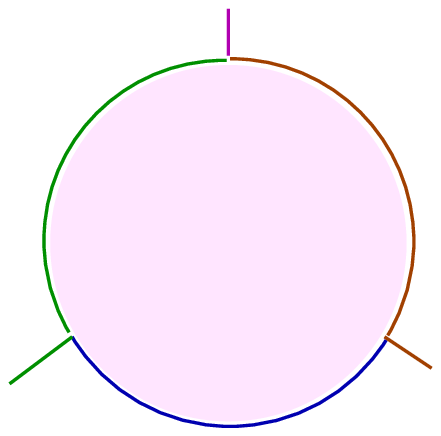}%
\end{picture}%
\setlength{\unitlength}{1973sp}%
\begingroup\makeatletter\ifx\SetFigFont\undefined%
\gdef\SetFigFont#1#2#3#4#5{%
  \reset@font\fontsize{#1}{#2pt}%
  \fontfamily{#3}\fontseries{#4}\fontshape{#5}%
  \selectfont}%
\fi\endgroup%
\begin{picture}(5348,4971)(76,-4444)
\put(4501,-3811){\makebox(0,0)[lb]{\smash{{\SetFigFont{12}{14.4}
{\familydefault}{\mddefault}{\updefault}$\chi_{ac}$}}}}
\put(2401,239){\makebox(0,0)[lb]{\smash{{\SetFigFont{12}{14.4}
{\familydefault}{\mddefault}{\updefault}$q_{cb}$}}}}
\put(4501,-1561){\makebox(0,0)[lb]{\smash{{\SetFigFont{12}{14.4}
{\familydefault}{\mddefault}{\updefault}$9_c$}}}}
\put(
76,-1411){\makebox(0,0)[lb]{\smash{{\SetFigFont{12}{14.4}
{\familydefault}{\mddefault}{\updefault}$9_b$}}}}
\put(2476,-4336){\makebox(0,0)[lb]{\smash{{\SetFigFont{12}{14.4}
{\familydefault}{\mddefault}{\updefault}$9_a$}}}}
\put(451,-3961){\makebox(0,0)[lb]{\smash{{\SetFigFont{12}{14.4}
{\familydefault}{\mddefault}{\updefault}$\chi_{ba}$}}}}
\end{picture}%
\end{center}
\caption{A disk diagram leading to a Yukawa coupling.}
\label{fig:yuk}
\end{figure}

In this situation, there are non-trivial interactions supported on disks whose
boundary is partly attached to three different branes, say of types $a$, $c$ and
$b$, provided the twist angles $\nu^{(i)}_{ba}$, $\nu^{(i)}_{ac}$, $\nu^{(i)}_{cb}$
for each $i$ are either the internal or the external angles of a triangle.
These interactions in the field-theory limit correspond to Yukawa couplings
between the fields of the chiral multiplets $q^{ac}$, $q^{cb}$ and $q^{ba}$,
like for instance the one associated to Fig. \ref{fig:yuk}:
\begin{equation}
 \label{yk}
\int d^4x \,\, Y_{acb}\, \Tr\big(\chi^{ac} q^{cb} \chi^{ba}\big)~,
\end{equation}
plus its supersymmetric completion terms. Altogether such interactions can be
encoded in the cubic superpotential
\begin{equation}
 \label{supyuk}
W_{\mathrm{Y}}=Y_{acb}\, \Tr\big(q^{ac} q^{cb} q^{ba}\big)~.
\end{equation}
If we rewrite the above superpotential in terms of the multiplets in the
supergravity basis via the rescalings (\ref{qQ}), we obtain
\begin{equation}
 \label{supyukQ}
W_{\mathrm{Y}}=Y_{acb}\, \big(
K_{ac}\,K_{bc}\,K_{ba}\big)^{\frac 12}\,\,\Tr\big(Q^{ac} Q^{cb} Q^{ba}\big)~.
\end{equation}
On the other hand, in the
effective supergravity action this superpotential must take the form
\begin{equation}
 \label{supyuksugra}
W_{\mathrm{Y}}= \ee^{K/2} \,{\widehat W}_{acb} \,\Tr\big(Q^{ac} Q^{cb}
Q^{ba}\big)~,
\end{equation}
where $\ee^{K/2}$ is the standard contribution of the bulk
K\"ahler potential and ${\widehat W}_{acb}$ are purely holomorphic functions
of the geometric moduli. Comparing these last two equations, we deduce that
\begin{equation}
\label{YW}
Y_{acb} = \ee^{K/2}\, \big(K_{ac}\,K_{cb}\,K_{ba}\big)^{-\frac 12}\,{\widehat W}_{acb}~.
\end{equation}
If we now use the bulk K\"ahler potential (\ref{kpot}) and the K\"ahler
metric $K_{ba}=\mathcal{Z}_{ba}\mathcal{X}_{ba}$, with $\mathcal{Z}_{ba}$
given in (\ref{KQ}) and similarly for $K_{ac}$ and $K_{cb}$, we
easily obtain
\begin{equation}
\label{YWus}
\begin{aligned}
Y_{acb} & = (4\pi)^{\frac 38} \,s_2^{-\frac 18}
\,\big(t^{(1)}_{2} t^{(2)}_{2} t^{(3)}_{2}\big)^{-\frac 18}
\,\big(u^{(1)}_{2} u^{(2)}_{2}u^{(3)}_{2}\big)^{\frac 14}\,
\big(\Gammab_{ac} \Gammab_{cb} \Gammab_{ba}\big)^{-\frac 14}\,
\big(\mathcal{X}_{ac} \mathcal{X}_{cb} \mathcal{X}_{ba}\big)^{-\frac 12}\,{\widehat W}_{acb} \\
& = \sqrt{4\pi}\,\,\ee^{\frac{\phi_4}{2}} \,\big(u^{(1)}_{2}
u^{(2)}_{2}u^{(3)}_{2}\big)^{\frac 14}\, \big(\Gammab_{ac}
\Gammab_{cb} \Gammab_{ba}\big)^{-\frac 14}\,
\big(\mathcal{X}_{ac} \mathcal{X}_{cb} \mathcal{X}_{ba}\big)^{-\frac 12}\, {\widehat W}_{acb}~,
\end{aligned}
\end{equation}
where $\phi_4=\phi_{10} -\frac{1}{2}\sum_i \log(T_2^{(i)})$
is the four-dimensional dilaton.

We now compare this finding with the results of Ref. \cite{Cremades:2004wa} for
the physical Yukawa couplings of toroidal models with magnetized D9 branes%
\footnote{In the T-dual intersecting brane version, these couplings have been
studied in Refs.~\cite{Cremades:2003qj,Cvetic:2003ch,Abel:2003vv,Lust:2004cx}.}.
The expression for $Y_{acb}$ is given in their Eq. (7.13). Setting to zero the
value of the Wilson lines, and rewriting it in terms of the supergravity moduli
through Eq.s (5.47)-(5.49) of the same reference, in our notation it reads
\begin{equation}
\label{cr}
Y_{acb} = \ee^{\frac{\phi_4}{2}}\, \big(u^{(1)}_{2}
u^{(2)}_{2}u^{(3)}_{2}\big)^{\frac 14}
\,\prod_{i=1}^3 \left| \frac{{\vartheta}_1^{(i)} {\vartheta}_2^{(i)}}{{\vartheta}_1^{(i)}
+ {\vartheta}_2^{(i)}}\right|^{\frac 14}
W'_{acb}~.
\end{equation}
The detailed expression of the quantities ${\vartheta}_1^{(i)}$, ${\vartheta}_2^{(i)}$
and $W'_{acb}$ is not relevant here; the only important
points that we want to emphasize are that $W'_{acb}$ is a holomorphic
function of the complex structure moduli $u^{(i)}$ and that the expression
(\ref{cr}) has been obtained starting from the non-abelian Yang-Mills theory on
the D9 branes, rather than from the full fledged DBI action. Therefore one
expects that it only represents the field theory limit of the string result%
\footnote{See Ref. \cite{Russo:2007tc} for a string theory calculation of the
Yukawa couplings and a direct derivation of the $\Gammab$ factors.}.
However, as already argued in Ref. \cite{Cremades:2004wa}, one can extend
\eq{cr} by observing that in the field theory limit $\alpha'\to 0$ ({\it i.e.}
in the small twist limit) one has
\begin{equation}
\label{ItoG}
\big(\Gammab_{ac} \Gammab_{cb} \Gammab_{ba}\big)^{-\frac
14}\,\sim\,\prod_{i=1}^3
\left| \frac{\vartheta_1^{(i)} \vartheta_2^{(i)}}{
\vartheta_1^{(i)} + \vartheta_2^{(i)}}\right|^{\frac 14}
~.
\end{equation}
With this understanding, \eq{cr} can then be generalized as
\begin{equation}
\label{crs}
Y_{acb} = \ee^{\frac{\phi_4}{2}}\, \big(u^{(1)}_{2}
u^{(2)}_{2}u^{(3)}_{2}\big)^{\frac 14}
\,\big(\Gammab_{ac} \Gammab_{cb} \Gammab_{ba}\big)^{-\frac 14}
\,W'_{acb}~,
\end{equation}
which agrees
with Eq. (\ref{YWus}) by taking $W'_{acb} =\sqrt{4\pi}\,\widehat W_{acb}$,
provided the non-holomorphic factors obey
\begin{equation}
\mathcal{X}_{ac} \mathcal{X}_{cb} \mathcal{X}_{ba}=1~.
\label{vincx}
\end{equation}
Indeed, stripping off the various factors of the K\"ahler potential
and of the K\"ahler metrics from the physical Yukawa couplings $Y_{acb}$
according to \eq{YWus}, we can obtain the expected holomorphic structure of
the superpotential only if (\ref{vincx}) is satisfied.

The simplest solution to this constraint is clearly
$\mathcal{X}_{ba}=\mathcal{X}_{cb}=\mathcal{X}_{ac}=1$. This would
imply that the K\"ahler metrics for the twisted chiral matter
fields are given by the expressions (\ref{KQ}) and (\ref{KQtilde}).
Notice that in this case the only dependence on the twist
parameters would be through the $\Gamma$-functions contained in the factors
$\big(\Gammab_{ba}\big)^{\frac12}$ and $\big(\Gammab_{ac}\big)^{\frac12}$ (see
Eq. (\ref{Gammab})). Such factors are the same as the ones that can be obtained
directly from a 3-point string scattering amplitude involving one (closed
string) geometric modulus and two twisted scalar fields, as explained in Refs.
\cite{Lust:2004cx,Bertolini:2005qh}. On the other hand, the
possibility of non-trivial $\mathcal{X}$ factors has been
considered in Refs. \cite{Akerblom:2007uc,Blumenhagen:2007ip}.
Besides satisfying the constraint (\ref{vincx}), such non-holomorphic factors
should also be related to a non-vanishing sigma-model anomaly term $\delta^{(0)}$,
as explained in Section \ref{subsec:hol_red}.
It would be very interesting to do an independent calculation to check this point.

We close with a few concluding remarks. Loop corrections to the
bulk K\"ahler potential or to the Einstein term in the bulk
action, which for Type II theories have been computed in Refs.
\cite{Antoniadis:1996vw,Berg:2005ja}, in general induce shifts of the supergravity
variables and in particular are responsible for a non-vanishing
$\delta$-term in Eq. (\ref{deltags}). However, such a $\delta$-term
appears to be of order $g_s \sim {\tilde g}_a^{\,2}$ and thus it does not
affect the form of the K\"ahler metric for the $\mathcal{N}=1$ twisted matter at tree-level.
Furthermore, using the known expressions for the K\"ahler potential and
K\"ahler metrics, in Ref. \cite{Billo:2007sw} we have explicitly checked that in
$\mathcal{N}=2$ SQCD no shift $\delta^{(0)}$ is produced. It would
be very interesting to explore further this issue and see whether and how
an anomalous term $\delta^{(0)}$ in $\mathcal{N}=1$ theories with magnetized branes
is possible.

\acknowledgments{We thank R. Russo for useful discussions.
This work is partially supported by the Italian MUR under contract
PRIN-2005023102 {``Strings, D-branes and Gauge Theories''}
and by the European Commission FP6 Programme under contract
MRTN-CT-2004-005104 ``{Constituents, Fundamental
Forces and Symmetries of the Universe}'', in which A.L. is associated to
University of Torino,
and R.M to INFN-Frascati.
We thank the Galileo Galilei Institute for Theoretical Physics for the
hospitality and the INFN for partial support during the completion of this
work.}

\appendix

\section{Technicalities}
\label{app:not}
In this Appendix we provide some technical details. In particular in Appendix
\ref{appsub:geom} we describe the background geometry and the
$\mathbb{Z}_2\times\mathrm{Z}_2$ orbifold action. In Appendix \ref{appsub:vert}
we write the vertex operators for the physical states in the various open string
sectors and their GSO properties, and finally in Appendix \ref{appsub:annuli} we
present the calculation of the mixed annulus amplitudes for the flavored
sectors.

\subsection{The background geometry}
\label{appsub:geom}
The metric and the $B$-field on the torus $\mathcal{T}_2^{(i)}$ are expressed
in terms of the complex moduli $T^{(i)}$ and of the K\"ahler moduli $T^{(i)}$
as follows:
\begin{equation}
\label{GB}
G^{(i)}= \frac{T_2^{(i)}}{U_2^{(i)}}\,
\begin{pmatrix}1 & U_1^{(i)} \\ U_1^{(i)} &
|U^{(i)}|^2\end{pmatrix}~~~~\mbox{and}~~~~
B^{(i)} = \begin{pmatrix}\,
0 & - T_1^{(i)} \\ T_1^{(i)} & 0
\end{pmatrix}~.
\end{equation}
With respect to the above metric, the orthonormal complex coordinates and
the corresponding string fields on the torus $\mathcal{T}_2^{(i)}$ are given by:
\begin{equation}
Z^i=\sqrt{\frac{T_2^{(i)}}{2U_2^{(i)}}}\left({X^{2i+2}+ U^{(i)}
X^{2i+3}}\right)~,~~~
\Psi^i=\sqrt{\frac{T_2^{(i)}}{2U_2^{(i)}}}\left({\psi^{2i+2}+
U^{(i)}\psi^{2i+3}}\right)
\label{zipsii}
\end{equation}
for $i=1,2,3$.

The (anti-chiral) spin-fields $S^{\dot{\mathcal{A}}}$ of the RNS formalism in
ten dimensions factorize in a product of four-dimensional and internal
spin-fields, according to
\begin{equation}
S^{\dot{\mathcal{A}}}
\to (S_\alpha S_{---},
S_\alpha S_{-++},S_\alpha S_{+-+},S_\alpha S_{++-},S^{\dot\alpha} S^{+++},
S^{\dot\alpha} S^{+--},S^{\dot\alpha} S^{-+-},S^{\dot\alpha} S^{--+})~,
\label{spin}
\end{equation}
where the index $\alpha$ ($\dot\alpha$) denotes positive (negative) chirality in
$\mathbb{R}^{1,3}$ and the labels $(\pm,\pm,\pm)$ on the internal spin-fields
denote charges $(\pm\frac12,\pm\frac12,\pm\frac12)$ under the three internal
$\mathrm{U}(1)$'s.

The orbifold group $\mathbb{Z}_2\times \mathbb{Z}_2$ contains three non-trivial
elements $h_i$ (subject to the relation $h_1 h_2=h_3$) acting on the internal
coordinates as follows
\begin{equation}
\label{frac2}
\begin{aligned}
h_1:~(Z^1,Z^2,Z^3) &\rightarrow (Z^1,-Z^2,-Z^3)~,\\
h_2:~(Z^1,Z^2,Z^3) &\rightarrow (-Z^1,Z^2,-Z^3)~,\\
h_3:~(Z^1,Z^2,Z^3) &\rightarrow (-Z^1,-Z^2,Z^3)~,
\end{aligned}
\end{equation}
and similarly for the $\Psi^{1,2,3}$ fields. We may summarize the transformation
properties (\ref{frac2}) for the conformal fields $\partial Z^i$ and $\Psi^i$
($i=1,2,3$) in the Neveu-Schwarz sector by means of the following table:
\begin{equation}
\label{ZPsirep}
\begin{tabular}{c|c}
conf. field & irrep \\
\hline
$\Big.\partial Z^i$, $\Psi^i$ & $R_i$
\end{tabular}~,
\end{equation}
where $\{R_I\}=\{R_0,R_i\}$ are the irreducible representations of
$\mathbb{Z}_2\times \mathbb{Z}_2$, identified by writing the character table of
the group
\begin{equation}
\label{frac3}
\begin{tabular}{c|cccc}
 & $e$ & $h_1$ & $h_2$ & $h_3$ \\
\hline
$R_0$ & 1 & ~1 & ~1 & ~1 \\
$R_1$ & 1 & ~1 & $-1$ & $-1$ \\
$R_2$ & 1 & $-1$ & ~1 & $-1$ \\
$R_3$ & 1 & $-1$ & $-1$ & ~1
\end{tabular}~.
\end{equation}
The Clebsh-Gordan series for these representations is simply given by
\begin{equation}
\label{frac4}
R_0\otimes R_I = R_I~~,~~~~
R_i\otimes R_j = \delta_{ij}  R_0 + |\epsilon_{ijk}| R_k~.
\end{equation}
The action of the orbifold group on  spin fields and spinor states
is given by a spinor representation of the geometrical rotations of $\pi$
in the various tori defined in \eq{frac2}. In particular, we choose it to be
given by
\begin{equation}
\label{orbspin}
\begin{aligned}
h_1& = \phantom{-~}\uno\otimes \sigma_3\otimes\sigma_3~,\\
h_2& = -\sigma_3\otimes \uno~\otimes\sigma_3~,\\
h_3& = -\sigma_3\otimes \sigma_3\otimes\uno~,\\
\end{aligned}
\end{equation}
which corresponds to the following table:
\begin{equation}
\label{spintransf}\begin{tabular}{c|c|c|c|c|c}
anti-chiral &  chiral  & $h_1$ & $h_2$ & $h_3$ & irrep \\
\hline
$\Big.S_{-++}$
& $S^{+--}$
& $~1$   & $~1$  & $~1$ & $R_0$ \\
$\Big.S_{---}$
& $S^{+++} $
& $~1$   & $-1$  & $-1$ & $R_1$ \\
$\Big.S_{++-}$
& $S^{--+} $
& $-1$  & $~1$  & $-1$ & $R_2$ \\
$\Big.S_{-+-} $
& $S^{+-+}$
& $-1$  & $-1$  & $1$ & $R_3$
\end{tabular}~.
\end{equation}
In particular, the only invariant spin fields are $S_{-++}$ and $S^{+--}$.

\subsection{The string vertices}
\label{appsub:vert}
We describe now in some more detail the string realization
of the massless $\mathcal{N}=1$ multiplets corresponding to the system of
magnetized branes we consider in the paper. This system is pictorially
represented in Fig.s \ref{fig:9a9b9c} and \ref{fig:5a9b9c}, where however only
the flavored multiplets and moduli respectively are indicated.

\paragraph{D$9_a$/D$9_a$ strings}
These strings are untwisted.
The NS massless vertices in the $(-1)$ superghost picture:
\begin{equation}
V_A(z) =(\pi\alpha')^{\frac12}\,{A_\mu}\,\psi^\mu(z)\,
\ee^{-\varphi(z)}\,\ee^{\ii p_\mu X^\mu(z)}~,
\label{vertA}
\end{equation}
and those of the  R sector, which we write in the $(-1/2)$ picture:
\begin{equation}
V_{\Lambda}(z)
=(2\pi\alpha')^{\frac34}\,\Lambda^{\alpha}\,S_{\alpha}(z)S_{-++}(z)\,
\ee^{-\frac12\varphi(z)}\,\ee^{\ii
p_\mu X^\mu(z)}
\label{vertlam}
\end{equation}
contain the d.o.f. of the $\mathcal{N}=1$ gauge multiplet. In these vertices,
the polarizations have canonical dimensions (this explains the dimensional
prefactors%
\footnote{See for example Ref. \cite{Billo:2002hm} for details on
the normalizations of vertex operators and scattering amplitudes.}) and are
$N_a\times N_a$ matrices transforming in the adjoint representation of
$\mathrm{SU}(N_a)$; here we neglect the $\mathrm{U}(1)$ factor associated to the
center of mass of the $N_a$ D9 branes. With respect to the orbifold group, the
D$9_a$/D$9_a$ strings carry Chan-Paton factors in the representation $R_0\times
R_0 = R_0$. Also the operator part of the vertices (\ref{vertA}) and
(\ref{vertlam}) must therefore be invariant under the orbifold. This is clearly
the case for $V_A$, and the gaugino vertex (\ref{vertlam}) contains the spin
field $S_{-++}$ which, according to (\ref{spintransf}) is indeed invariant.

\paragraph{D$9_b$/D$9_a$ and D$9_a$/D$9_b$ strings}
Next we consider the strings stretching between two different stacks of branes.
For definiteness we focus on those connected on one end
to the color D$9_a$ branes, and the other end to the flavor
D$9_b$ branes. These strings have Chan-Paton factors which, with respect to the
orbifold group, belong to the representation
\begin{equation}
 \label{CPR1}
R_1\otimes R_0 = R_1~.
\end{equation}

To write the vertex operators it is convenient to introduce the following
notation
\begin{equation}
\sigma(z)\equiv\prod_{i=1}^3\sigma_{\nu_{ba}^{(i)}}(z)\quad,\quad
s(z)\equiv\prod_{i=1}^3 S_{\nu_{ba}^{(i)}}(z)~,
\label{sigma}
\end{equation}
where $\sigma_{\nu_{ba}^{(i)}}$ and $S_{\nu_{ba}^{(i)}}$ are respectively
the bosonic and fermionic twist fields in the
$i$-th torus whose conformal dimensions are
\begin{equation}
h_{\sigma}^{(i)}= \frac{1}{2}\nu_{ba}^{(i)}\big(1-\nu_{ba}^{(i)}\big)
~~~{\mbox{and}}~~~h_{S}^{(i)}= \frac{1}{2}\big(\nu_{ba}^{(i)}\big)^2~.
\label{h}
\end{equation}
Then, the physical massless state in the NS sector of the D$9_b$/D$9_a$ is a
complex scalar described by the following vertex operator:
\begin{equation}
V_{q}(z) = (2\pi\alpha')^{\frac12}\,q\,\,
\sigma(z):\!\bar\Psi^1(z)s(z)\!:
{\rm e}^{-\varphi(z)}\,{\rm e}^{{\rm i}p_\mu X^\mu(z)}~,
\label{vertn1scal}
\end{equation}
which can be easily checked%
\footnote{Remember that the conformal dimension of ${\rm e}^{-\varphi(z)}$ and
${\rm e}^{-\frac12\varphi(z)}$ is respectively equal to $\frac{1}{2}$ and
$\frac{3}{8}$.} to have conformal dimension 1 for $p^2=0$ when the twists
satisfy the supersymmetry condition $\nu_{ba}^{(1)}= \nu_{ba}^{(2)} +
\nu_{ba}^{(3)} $. The operator $s(z)$ is invariant under the orbifold, so that
$:\!\bar\Psi^1(z) s(z)\!:$ belongs to the representation $R_1$ and compensates
the non-trivial transformation of the Chan-Paton factors (see Eq. (\ref{CPR1})).

In the R sector the massless states are described by the following
vertex operator:
\begin{equation}
V_{\chi}(z) = (2\pi\alpha')^{\frac34}\,
{\chi}^{\alpha}\,S_{\alpha}(z)\,\sigma(z)\Sigma(z)
\,
{\rm e}^{-\frac12\varphi(z)}\,{\rm e}^{{\rm i}p_\mu X^\mu(z)}~,
\label{vertn1ferm}
\end{equation}
where
\begin{equation}
\Sigma(z)=\prod_{i=1}^3 S_{(\nu_{ba}^{(i)}-\frac12)}(z)~.
\label{Sigma}
\end{equation}
Again, one can easily check that this vertex operator has
conformal dimension 1 for $p^2=0$ when the twists satisfy the supersymmetry
condition. With respect to the orbifold group, the operator $\Sigma(z)$
transforms in the representation $R_1$, just as the spin field $S_{---}$ to
which it reduces in the limit of vanishing twists (see Eq. (\ref{spintransf})).
The chirality of the spinor $\chi$ is fixed by the GSO projection,
with
\begin{equation}
\label{GSOba}
P_{\mathrm{GSO}}^{9b/9a}= \frac{1 + (-)^F}{2}~,
\end{equation}
and the fact that the vertex (\ref{vertn1ferm}) survives the projection means that
on the corresponding state we have
\begin{equation}
 \label{GSOR}
(-)^F \ket{S_{\alpha}\sigma\Sigma} = + \ket{S_{\alpha}\sigma\Sigma}~.
\end{equation}
The vertices (\ref{vertn1scal}) and (\ref{vertn1ferm}) belong to the $\bar N_a$
representation of the color group,
and have a flavor degeneracy described in  the main text (see Eqs. (\ref{Nab})
and (\ref{iab})).

If we consider the D$9_a$/D$9_b$ strings we find the conjugate
scalar $q^\dagger$:
\begin{equation}
V_{q^\dagger}(z) = (2\pi\alpha')^{\frac12}\,q^\dagger\,\,
\bar\sigma(z):\!\Psi^1(z)\bar s(z)\!:
{\rm e}^{-\varphi(z)}\,{\rm e}^{{\rm i}p_\mu X^\mu(z)}
\label{vertn1scald}
\end{equation}
and the conjugate fermion $\chi^\dagger$:
\begin{equation}
V_{\chi^\dagger}(z) = (2\pi\alpha')^{\frac34}\,
{\chi}_{\dot\alpha}\,S^{\dot\alpha}(z)\,\bar\sigma(z)\bar\Sigma(z)
\,
{\rm e}^{-\frac12\varphi(z)}\,{\rm e}^{{\rm i}p_\mu X^\mu(z)}~,
\label{vertn1fermd}
\end{equation}
where the anti-twist fields are defined as follows:
\begin{equation}
\bar\sigma(z)\equiv\prod_{i=1}^3\sigma_{1-\nu_{ba}^{(i)}}(z)~,\quad
\bar s(z)\equiv\prod_{i=1}^3 S_{-\nu_{ba}^{(i)}}(z)
~,\quad
\bar\Sigma(z)\equiv\prod_{i=1}^3 S_{\frac 12 -\nu_{ba}^{(i)}}~.
\label{asigma}
\end{equation}
These vertices transform in the fundamental representation $N_a$ of
the color group and are degenerate in flavor as the $q$ and $\chi$
vertices.

Let us notice that the operator appearing in \eq{vertn1fermd} is not the CFT
conjugate of the one appearing in \eq{vertn1ferm}, which would be given%
\footnote{In fact, the nontrivial
overlap between the 4d spin fields is $\braket{S_\beta}{S_\alpha} \propto
\epsilon_{\alpha\beta}$.}
by $S_\beta \bar\sigma \bar\Sigma$.
This latter has the same $F$-parity as in \eq{GSOR}, so that the state
corresponding to the vertex (\ref{vertn1fermd}) must have the opposite one:
\begin{equation}
 \label{GSORa}
(-)^F \ket{S_{\dot\alpha}\bar\sigma\bar\Sigma} = -
\ket{S_{\dot\alpha}\bar\sigma\bar\Sigma}~.
\end{equation}
The anti-chiral vertex (\ref{vertn1fermd}) is selected because in the R sector of
the D$9_a$/D$9_b$ strings we take
\begin{equation}
\label{GSOab}
P_{\mathrm{GSO}}^{9a/9b}= \frac{1 - (-)^F}{2}~,
\end{equation}
as opposed to \eq{GSOba}. This has important consequences for the annulus
partition function of the flavored strings; see the discussion of this point
after Eq. (4.34) of Ref. \cite{Billo:2007sw}, in an $\mathcal{N}=2$ context.

\paragraph{D$9_a$/D$9_c$ and D$9_c$/D$9_a$ strings}
The open strings stretching between the color D$9_a$ branes and the flavor
branes of type D$9_c$ have essentially the same characteristics of the
D$9_b$/D$9_a$ and D$9_a$/D$9_b$ ones, provided one replaces the twist angles
$\nu^{(i)}_{ba}$ with
\begin{equation}
\label{nuac}
\nu_{ac}^{(i)} = \nu_{a}^{(i)} - \nu_{c}^{(i)}~.
\end{equation}

In the NS sector of the D$9_a$/D$9_c$ strings we find a complex scalar $\tilde
q$ associated to the massless vertex
\begin{equation}
V_{\tilde q}(z) = (2\pi\alpha')^{\frac12}\,\tilde q\,\,
\tilde \sigma(z):\!\bar\Psi^1(z)\tilde s(z)\!:
{\rm e}^{-\varphi(z)}\,{\rm e}^{{\rm i}p_\mu X^\mu(z)}~,
\label{vertn1scalt}
\end{equation}
while in the R sector we have a chiral spinor:
\begin{equation}
V_{\tilde \chi}(z) = (2\pi\alpha')^{\frac34}\,
\tilde{\chi}^{\alpha}\,S_{\alpha}(z)\,\tilde\sigma(z)\tilde\Sigma(z)
\,
{\rm e}^{-\frac12\varphi(z)}\,{\rm e}^{{\rm i}p_\mu X^\mu(z)}~.
\label{vertn1fermt}
\end{equation}
Here the twist/spin fields $\tilde\sigma$, $\tilde s$ and $\tilde \Sigma$ are
defined analogously to Eq.s (\ref{sigma}) and (\ref{Sigma}) in terms of the
angles $\nu^{(i)}_{ac}$ of \eq{nuac}. These vertices transform in the
{fundamental} representation $N_a$ of the color group and are degenerate in
flavor as indicated in \eq{Nac} in the main text. Notice that for the $(a,c) $
``intersection'' the r\^ole of the color and flavor branes is switched, with
respect to the  $(b,a)$ intersection, for what concerns the GSO/chirality
projection.

\paragraph{E$5_a$/E$5_a$ strings}
The polarizations in all vertices arising from these strings are $k\times k$
matrices transforming in the adjoint of $\mathrm{U}(k)$, and belong to the
trivial representation of the orbifold group. The operator part of the vertices
must therefore also be invariant under the orbifold.

In the NS sector we have the vertices
\begin{subequations}
\label{vertNS}
\begin{align}
\label{verta}
&V_{a}(z)= g_{5_a}\,(2\pi\alpha')^{\frac12}\,a_\mu
\,\psi^{\mu}(z)\,\ee^{-\varphi(z)}~,
\\
\label{vertd} &V_D(z)= {D_c}\,(\pi\alpha')^{\frac12}\,\bar\eta_{\mu\nu}^c\,
\psi^\nu(z)\psi^\mu(z)~,
\end{align}
\end{subequations}
where $\bar\eta^c_{\mu\nu}$ are the three anti-self-dual 't Hooft
symbols. In the R sector, we have
\begin{equation}
\label{vertR1}
V_{M}(z) = \frac{g_{5_a}}{\sqrt 2}\,(2\pi\alpha')^{\frac34}\,
{M}^{\alpha}\,S_{\alpha}(z) S_{-++}(z)\,\ee^{-\frac{1}{2}\varphi(z)}~,\\
\end{equation}
\begin{equation}
\label{vertR2}
V_{\lambda}(z) =
{{\lambda_{\dot\alpha}}}\,(2\pi\alpha')^{\frac34}\,S^{\dot\alpha}(z)S^{+--}(z)
\,\ee^{-\frac{1}{2}\varphi(z)}~.
\end{equation}

\paragraph{D$9_a$/E$5_a$ and E$5_a$/D$9_a$ strings}
In the NS sector  of D$9_a$/E$5_a$ strings we have the vertices
\begin{equation}
\label{vertw}
V_w(z)= \frac{g_{5_a}}{\sqrt{2}}\,(2\pi\alpha')^{\frac12}\,{w}_{\dot\alpha}\,
\Delta(z) S^{\dot\alpha}(z)\,\ee^{-\varphi(z)}~.
\end{equation}
Here $\Delta$ is the twist operator with conformal weight $1/4$ which changes
the boundary conditions of the non-compact coordinates $X^\mu$ from Neumann to
Dirichlet.

In the R sector there is the vertex
\begin{equation}
V_{\mu}(z) = \frac{g_{5_a}}{\sqrt{2}}\,
(2\pi\alpha')^{\frac34}\,{\mu}\, \Delta(z) S_{-++}(z)\,
\ee^{-{\frac12}\varphi(z)}~.
\label{vertmu}
\end{equation}
Both in (\ref{vertw}) and (\ref{vertmu}) the polarizations are $N_a\times k$
matrices which transform in the bi-fundamental representation ${N}_a\times \bar
k$ of ${{\mathrm U}(N_a)\times\mathrm U}(k)$. The Chan-Paton factors and the
operator part of these vertices are invariant under the orbifold group.

The charged moduli associated to the E$5_a$/D$9_a$ strings, denoted by $\bar
w_{\dot\alpha}$ and ${\bar\mu}$, transform in the $\bar{N}_a\times  k$
representation and are described by vertex operators of the same form as
(\ref{vertw}) and (\ref{vertmu}) with $\Delta(z)$ replaced by the anti-twist
$\bar\Delta(z)$, corresponding to DN (instead of ND) boundary conditions along
the space-time directions. In particular, we have
\begin{equation}
V_{\bar\mu}(z) = \frac{g_{5_a}}{\sqrt{2}}\,
(2\pi\alpha')^{\frac34}\,{\bar\mu}\, \bar\Delta(z) S_{-++}(z)\,
\ee^{-{\frac12}\varphi(z)}~.
\label{vertmubar}
\end{equation}
Notice that this vertex is not the CFT conjugate of \eq{vertmu}, which would
contain the operator $\bar\Delta S^{+--}$.

\paragraph{D$9_b$/E$5_a$  and E$5_a$/D$9_b$ strings}
The Chan-Paton factors for these strings transform, analogously to \eq{CPR1}, in
the representation $R_1$ of the orbifold group.

Since there is no momentum available for these strings, it is not possible to
construct physical vertices in the NS sector. The only physical vertex for
D$9_b$/E$5_a$ strings is in the R sector:
\begin{equation}
\label{vertmup}
V_{\mu'}(z)= \frac{g_{5_a}}{\sqrt{2}}\,(2\pi\alpha')^{\frac34}\,
\mu'\, \Delta(z) \, \sigma(z)\, \Sigma(z)
\,
\ee^{-{\frac12}\varphi(z)}~.
\end{equation}
Notice that the operator part of this vertex transforms in the $R_1$
representation (see the discussion after \eq{Sigma} above), and this
appropriately compensates the transformation of the Chan-Paton factors.
The polarization transforms in the anti-fundamental of $\mathrm{U}(k)$ and
carries a flavor degeneracy as in \eq{Nab}.
The vertex (\ref{vertmup}) survives the GSO projection
\begin{equation}
\label{GSO9b5a}
P_{\mathrm{GSO}}^{9b/5a}= \frac{1 + (-)^F}{2}
\end{equation}
hence, on the corresponding state, we have
\begin{equation}
 \label{GSORa5}
(-)^F \ket{\Delta \sigma \Sigma} = +
\ket{\Delta \sigma \Sigma}~.
\end{equation}

If we consider now the mixed strings with the opposite orientation, the
E$5_a$/D$9_b$ ones, we can only construct a massless vertex using the operator
$\bar\Delta\bar\sigma\bar\Sigma$. This operator is the CFT conjugate of the
operator appearing in \eq{vertmup}, and has therefore the same $F$-parity.
However, in the R sector, as usual we have to take the opposite projection with
respect to \eq{GSO9b5a}, namely
\begin{equation}
\label{GSO5a9b}
P_{\mathrm{GSO}}^{5a/9b}= \frac{1 - (-)^F}{2}~.
\end{equation}
Therefore, no physical state with this orientation survives the projection.

\paragraph{E$5_a$/D$9_c$ and D$9_c$/E$5_a$ strings}
In this sector, we find a physical vertex in the R sector of the E$5_a$/D$9_c$
given by
\begin{equation}
\label{vertmupt}
V_{{\tilde\mu}'}(z)= \frac{g_{5_a}}{\sqrt{2}}\,(2\pi\alpha')^{\frac34}\,
{\tilde\mu}'\, \Delta(z) \, \tilde\sigma(z)\, \tilde\Sigma(z)
\,
\ee^{-{\frac12}\varphi(z)}~.
\end{equation}
The polarization transforms in the {fundamental} of $\mathrm{U}(k)$ and
carries a flavor degeneracy as in \eq{Nac}.

In the mixed strings   D$9_c$/E$5_a$ with the opposite orientation, no physical
state survives the GSO projection.

\subsection{The flavored annuli}
\label{appsub:annuli}
In this section we compute in more detail the contribution of E$5_a$/D$9_b$ and
D$9_b$/E$5_a$ strings to the annulus amplitude, deriving the expressions given
in Eq.s (\ref{z59bis}) and (\ref{59tw}) for the one-loop traces
${\cal{A}}_{5_a;9_b}^{h_I}$ containing the insertions of the various elements of
the orbifold group.

The untwisted sector corresponds to the insertion the identity element
$h_0\equiv e$. For the even spin structures, one finds that contributions
corresponding to the two possible orientations are equal. Adding them, one finds
\begin{equation}
\begin{aligned}
{\cal{A}}_{5_a;9_b}^{e}  & = \frac{\ii\, k\,{N_F}}{2\pi}
\int_{0}^{\infty} \frac{d\tau}{2 \tau}\!\!
\sum_{\substack{\alpha,\beta=0\\(\alpha,\beta)\neq(1,1)}}^1
\!\!(-1)^{\alpha+\beta+\alpha\beta} ~~
\frac{\theta \ssb \alpha\beta (0 | \ii \tau)}{\theta^\prime_{1}(0|\ii \tau)} \\
 & \times \left( \frac{\theta \ssb \alpha\beta( \frac{\ii \tau}{2} | \ii
\tau)\,\theta^\prime_{1} (0|\ii \tau)}{\theta \ssb \alpha\beta \left( 0 | \ii \tau\right)
\theta_{1} ( \frac{\ii \tau}{2}|\ii\tau)}\right)^2
\prod_{i=1}^{3}\frac{\theta \ssb \alpha\beta (\ii \nu^{(i)}\tau|\ii\tau)}{\theta_1
(\ii\nu^{(i)}\tau|\ii\tau)}
\end{aligned}
\label{d5d9b}
\end{equation}
where $N_F$ is given in Eq. (\ref{nf}) and we have used the following
notations%
\footnote{For the $\theta$-functions we use the notations listed in
Appendix A of Ref.~\cite{Di Vecchia:2005vm} where also the Riemann identity,
that will be used several times in the following, is given.}
\begin{eqnarray}
\theta\ssb 00 \equiv  \theta_3~~;~~
\theta\ssb 01 \equiv  \theta_4~~;~~
\theta\ssb 10 \equiv  \theta_2~~;~~
\theta\ssb 11 \equiv  \theta_1~.
\label{notat}
\end{eqnarray}
Moreover, we have denoted with primes the derivatives of the $\theta$-functions
with respect to their first argument, {\it i.e.}
\begin{equation}
 \label{dertheta}
\theta^\prime_{1}(\nu|\ii \tau) \equiv \partial_z
\theta_1(z|\ii\tau)\big|_{z=\nu}~.
\end{equation}

The contribution of the odd spin structure R$(-1)^F$ for a given orientation is
actually infinite and cannot be regularized using the procedure described in
Ref.~\cite{Billo:1998vr}. However, because of the different GSO projection to be
employed in the two oriented sectors (see Appendix \ref{appsub:vert}), one gets
a complete cancellation between the two orientations.

Eq. (\ref{d5d9b}), which coincides with Eq. (2.11) of Ref.~\cite{Abel:2006yk},
can be recast in a simpler form by first using the following relation
\begin{equation}
\left( \frac{\theta \ssb \alpha\beta(\frac{\ii \tau}{2}|\ii\tau)\,\,
\theta^\prime_{1}(0|\ii \tau)}{\theta \ssb \alpha\beta(0|\ii\tau)\,\,
\theta_{1}(\frac{\ii \tau}{2}|\ii\tau)}\right)^2 =
\frac{\theta^{\prime\prime}\ssb \alpha\beta(0|\ii\tau)}{\theta\ssb \alpha\beta(0|\ii\tau)}
- \partial_{z}^{2} \log \theta_1 (z|\ii\tau)\big|_{z=
\frac{\ii\tau}{2}}~,
\label{equa45}
\end{equation}
and then the Riemann identity
\begin{equation}
\begin{aligned}
& \sum_{\alpha,\beta=0 }^1 (-1)^{\alpha+\beta+\alpha\beta} ~~
  \theta\ssb \alpha\beta(z|\ii\tau)
\prod_{i=1}^{3} \theta\ssb \alpha\beta(\ii\tau\nu^{(i)}|\ii\tau)
\\
& = - \,2\, \theta_1\big(\frac{z}{2} | \ii \tau \big)\,
  \theta_1 \big(-\frac{z}{2} + \ii\tau \nu^{(1)}| \ii \tau \big)\,
  \theta_1 \big( \frac{z}{2} + \ii\tau \nu^{(2)}| \ii \tau \big)\,
  \theta_1 \big( \frac{z}{2} + \ii\tau \nu^{(3)}| \ii \tau \big)~.
\end{aligned}
\label{riema76}
\end{equation}
The last term of Eq. (\ref{equa45}) does not contribute when this is inserted
into Eq. (\ref{d5d9b}) since the right hand side of Eq. (\ref{riema76}) is zero
for $\nu=0$. Therefore we get
\begin{equation}
{\cal{A}}_{5_a;9_b}^{e}   =  \frac{\ii\, k\,{N_F}}{2\pi}
\int_{0}^{\infty} \frac{d\tau}{2\tau}
\!\!  \sum_{\substack{\alpha,\beta=0\\(\alpha,\beta)\neq(1,1)}}^1
\!\!\!(-1)^{\alpha+\beta+\alpha\beta}\,\,
\frac{\theta^{\prime\prime}\ssb
\alpha\beta(0|\ii\tau)}{\theta^\prime_{1}(0|\ii \tau)}\,
\prod_{i=1}^{3}\frac{\theta\ssb \alpha\beta(\ii\nu^{(i)}\tau|\ii\tau) }{\theta_1
(\ii\nu^{(i)}\tau|\ii\tau)}~.
\label{d5d9d}
\end{equation}
Taking the second derivative  with respect to $z$ of the Riemann identity
(\ref{riema76}) and evaluating it at $z=0$, we obtain another identity which
allows us to rewrite Eq. (\ref{d5d9d}) as follows:
\begin{eqnarray}
{\cal{A}}_{5_a;9_b}^{e} & = &\frac{\ii\, k\,{N_F}}{2 \pi}
\int_{0}^{\infty} \frac{d\tau}{2 \tau}
\label{z59bis2}\\
& &\times\,
\Big[ \partial_{z}\log \theta_1 ( z | \ii \tau )\big|_{z = \ii \tau \nu^{(1)}}
-\partial_{z}\log \theta_1 ( z | \ii \tau )\big|_{z = \ii \tau \nu^{(2)}}
-\partial_{z}\log \theta_1 ( z | \ii \tau )\big|_{z = \ii \tau \nu^{(3)}}
\Big]~.
\nonumber
\end{eqnarray}
This is precisely Eq. (\ref{z59bis}) of the main text.

Let us now turn to the three amplitudes ${\cal A}_{5_a;9_b}^{h_i}$ containing
the insertion on the non-trivial elements $h_i$ of the orbifold group. Again the
(divergent) contribution of the odd spin structure R$(-1)^F$ cancels when we sum
over the two orientations of the flavored strings, while the even spin
structures give the following contribution
\begin{eqnarray}
{\cal A}_{5_a;9_b}^{h_i} & = &\frac{\ii\, k\,{N_F}}{2\pi}
\int_0^\infty \frac{d\tau}{2\tau}\,\frac{1}{\theta^\prime_{1}(0|\ii
\tau)}
\!\!\!
\sum_{\substack{\alpha,\beta=0\\(\alpha,\beta)\neq(1,1)}}^1
\!\!\!(-1)^{\beta+\alpha\beta} \,\, [R_1(h_i)]^\alpha
\label{59}
\\
&&\left( \theta^{\prime\prime}\ssb \alpha\beta(0|\ii\tau) -
\theta\ssb \alpha\beta(0|i\tau)\,
\partial_{z}^2\log\theta_1(z|\ii\tau)\big|_{z=\frac{\ii\tau}{2}}\right)
\frac{\theta\ssb
\alpha\beta(\ii\tau\nu^{(i)}|\ii\tau)}{\theta_1(\ii\tau\nu^{(i)}|i \tau)}
\,\prod_{j\neq i=1}^3\frac{
\theta\ssb{\alpha}{\beta+1}(\ii\tau\nu^{(j)}|\ii\tau)}{
\theta_2(\ii\tau\nu^{(j)}|\ii\tau)}~,
\nonumber
\end{eqnarray}
where in the second line we have already used the relation (\ref{equa45}).
To proceed we have to distinguish the two cases corresponding to $h_1$, for
which
$R_1(h_1)=1$, and to $h_2, h_3$, for which $R_1(h_2) = R_1(h_3)=-1$.
In the first case Eq. (\ref{59}) becomes
\begin{equation}
\begin{aligned}
{\cal A}_{5_a;9_b}^{h_1} = & \,\frac{\ii\, k\,{N_F}}{2\pi}
\int_0^\infty \frac{d\tau}{2\tau}\frac{1}{\theta^\prime_{1}(0|\ii
\tau)}
\!\!\!
\sum_{\substack{\alpha,\beta=0\\(\alpha,\beta)\neq(1,1)}}^1
(-1)^{\alpha+\beta+\alpha\beta}
\\
& \times \left( \theta^{\prime\prime}\ssb \alpha\beta(0|\ii\tau) -
\theta\ssb \alpha\beta(0|i\tau)\,
\partial_{z}^2\log\theta_1(z|\ii\tau)\big|_{z=\frac{\ii\tau}{2}}\right)
\\
& \times
\frac{\theta\ssb \alpha\beta(\ii\tau\nu^{(1)}|\ii\tau)}
{\theta_1(\ii\tau\nu^{(1)}|\ii\tau)}\,\,
\frac{\theta\ssb{\alpha}{\beta+1}(\ii\tau\nu^{(2)}|\ii\tau)}
{\theta_2(\ii\tau\nu^{(2) }|\ii\tau)}\,\,
\frac{\theta\ssb{\alpha}{\beta-1}(\ii\tau\nu^{(3)}|\ii\tau)}
{\theta_2(\ii\tau\nu^{(3) }|\ii\tau)}~,
\end{aligned}
\label{59h1}
\end{equation}
where we have used the fact that
\begin{equation}
\theta\ssb{\alpha}{\beta-1}(z | \ii \tau) = (-1)^\alpha
\theta\ssb{\alpha}{\beta+1}(z | \ii \tau)~.
\label{b+b-}
\end{equation}
We can then exploit the identity
\begin{equation}
\begin{aligned}
\sum_{\alpha,\beta=0 }^1 &(-1)^{\alpha+\beta+\alpha\beta} \,
\theta\ssb \alpha\beta(z| \ii \tau)\,
\theta\ssb \alpha\beta(\ii \tau \nu^{(1)} | \ii \tau )\,
\theta\ssb{\alpha}{\beta+1}(\ii \tau \nu^{(2)} | \ii \tau )\,
\theta\ssb{\alpha}{\beta-1}(\ii \tau \nu^{(3)} | \ii \tau )
\\
= \,&\, 2 \,\theta_1 \big(\frac{z}{2} |\ii \tau \big)
\,\theta_1 \big( -\frac{z}{2} + \ii \tau \nu^{(1)} | \ii \tau \big)
\,\theta_2 \big( \frac{z}{2} + \ii \tau \nu^{(2)} | \ii \tau \big)
\,\theta_2 \big( \frac{z}{2} + \ii \tau \nu^{(3)} | \ii \tau \big)
\label{mainid}
\end{aligned}
\end{equation}
to check that the term containing the second derivative of the logarithm of
$\theta_1$ in Eq. (\ref{59h1}) does not give any contribution. The remaining
terms can be computed by differentiating twice with respect to $z$ both sides of
Eq. (\ref{mainid}) and putting $z=0$. In this way we get
\begin{eqnarray}
{\cal{A}}_{5_a;9_b}^{h_1} & = &\frac{\ii\, k\,{N_F}}{2 \pi}
\int_{0}^{\infty} \frac{d\tau}{2 \tau}
\label{h1fi}\\
& &\times\,
\Big[ \partial_{z}\log \theta_1 ( z | \ii \tau )\big|_{z = \ii \tau \nu^{(1)}}
-\partial_{z}\log \theta_2 ( z | \ii \tau )\big|_{z = \ii \tau \nu^{(2)}}
-\partial_{z}\log \theta_2 ( z | \ii \tau )\big|_{z = \ii \tau \nu^{(3)}}
\Big]~.
\nonumber
\end{eqnarray}

Let us now consider the amplitude with $h_2$ inserted. In this case
Eq. (\ref{59}) becomes
\begin{equation}
\begin{aligned}
{\cal A}_{5_a;9_b}^{h_2}  = & \,\frac{\ii\, k\,{N_F}}{2\pi}
\int_0^\infty \frac{d\tau}{2\tau}\frac{1}{\theta^\prime_{1}(0|\ii
\tau)}
\!\!\!
\sum_{\substack{\alpha,\beta=0\\(\alpha,\beta)\neq(1,1)}}^1
(-1)^{\alpha+\beta+\alpha\beta}
\\
& \times \left( \theta^{\prime\prime}\ssb{\alpha}{\beta-1}(0|\ii\tau) -
\theta\ssb{\alpha}{\beta-1}(0|\ii\tau)\,
\partial_{z}^2\log\theta_1(z|\ii\tau)|_{z=\frac{\ii\tau}{2}}\right)
\\
& \times
\frac{\theta\ssb{\alpha}{\beta+1}(i\tau\nu^{(2)}|\ii\tau)}
{\theta_1(i\tau\nu^{2}|i\tau)}\,\,
\frac{\theta\ssb \alpha\beta(\ii\tau\nu^{(3)}|\ii\tau)}
{\theta_2(\ii\tau\nu^{(3)}|\ii\tau)}\,\,
\frac{\theta\ssb \alpha\beta(\ii\tau\nu^{(1)}|\ii\tau)}
{\theta_2(\ii\tau\nu^{(1)}|\ii\tau)}
\end{aligned}
\label{59h2}
\end{equation}
after making the substitution $\beta+1 \rightarrow \beta$ and using Eq.
(\ref{b+b-}). Then, by exploiting the following relation:
\begin{eqnarray}
\sum_{\scriptstyle  \alpha,\beta=0 }^1 (-1)^{\alpha+\beta+\alpha\beta} &&\,
\theta\ssb{\alpha}{\beta-1}(z |\ii\tau)\,
\theta\ssb{\alpha}{\beta+1}(\ii\tau \nu^{(2)} |\ii\tau )\,
\theta\ssb \alpha\beta (\ii\tau\nu^{(3)} |\ii\tau )\,
\theta\ssb \alpha\beta(\ii\tau\nu^{(1)} |\ii\tau )\label{mainidn}
\\
&& =\, 2\,  \theta_1 \big(\frac{z}{2}|\ii\tau \big)
\,\theta_1 \big(\frac{z}{2} + \ii\tau\nu^{(2)}|\ii\tau \big)
\,\theta_2 \big(\frac{z}{2} + \ii\tau\nu^{(3)}|\ii\tau \big)
\,\theta_2 \big(-\frac{z}{2} +\ii\tau\nu^{(1)}|\ii\tau \big)~,
\nonumber
\end{eqnarray}
we can show that the second term in the second line in
Eq. (\ref{59h2}) is zero. The first term can instead be computed by
differentiating twice both sides of Eq. (\ref{mainidn}), and we
get
\begin{eqnarray}
{\cal{A}}_{5_a;9_b}^{h_2} & = &\frac{\ii\, k\,{N_F}}{2 \pi}
\int_{0}^{\infty} \frac{d\tau}{2 \tau}
\label{h2fi}\\
& &\times\,
\Big[ \partial_{z}\log \theta_1 ( z | \ii \tau )\big|_{z = \ii \tau \nu^{(2)}}
-\partial_{z}\log \theta_2 ( z | \ii \tau )\big|_{z = \ii \tau \nu^{(3)}}
-\partial_{z}\log \theta_2 ( z | \ii \tau )\big|_{z = \ii \tau \nu^{(1)}}
\Big]~.
\nonumber
\end{eqnarray}

Finally the amplitude with the insertion of $h_3$ is obtained from the previous
expression by the exchange $(i=2) \leftrightarrow (i=3)$, namely
\begin{eqnarray}
{\cal{A}}_{5_a;9_b}^{h_3} & = &\frac{\ii\, k\,{N_F}}{2 \pi}
\int_{0}^{\infty} \frac{d\tau}{2 \tau}
\label{h3fi}\\
& &\times\,
\Big[ \partial_{z}\log \theta_1 ( z | \ii \tau )\big|_{z = \ii \tau \nu^{(3)}}
-\partial_{z}\log \theta_2 ( z | \ii \tau )\big|_{z = \ii \tau \nu^{(2)}}
-\partial_{z}\log \theta_2 ( z | \ii \tau )\big|_{z = \ii \tau \nu^{(1)}}
\Big]~.
\nonumber
\end{eqnarray}
Eqs. (\ref{h1fi}), (\ref{h2fi}) and (\ref{h3fi}) can be written in
the compact form reported in Eq. (\ref{59tw}) of the main text.

\end{document}